\documentclass[sigconf,balance=false]{acmart}
\usepackage{popets}
\setcopyright{popets}
\copyrightyear{2024}

\acmYear{2024}
\acmVolume{2024}
\acmNumber{4}

\acmDOI{10.56553/popets-2024-0001}

\acmISBN{}
\acmConference{Proceedings on Privacy Enhancing Technologies}
\usepackage{amsmath}
\usepackage{pifont}
\usepackage{xcolor}
\newcommand{\cmark}{\ding{51}}
\newcommand{\xmark}{\ding{55}}
\usepackage{adjustbox}
\usepackage{hyperref}
\usepackage{array}
\newcolumntype{R}[2]{%
    >{\adjustbox{angle=#1,lap=\width-(#2)}\bgroup}%
    l%
    <{\egroup}%
}

\newcommand*\rotf{\multicolumn{1}{R{90}{.5em}}}

\begin{document}

\title{Fantômas: Understanding Face Anonymization Reversibility}

\author{Julian Todt}
\affiliation{%
 \department{KASTEL Security Research Labs}
 \institution{Karlsruhe Institute of Technology}
 \streetaddress{Am Fasanengarten 5}
 \city{Karlsruhe}
 \state{}
 \country{Germany}
 \postcode{76131}
}
\email{julian.todt@kit.edu}

\author{Simon Hanisch}
\affiliation{%
 \department{Centre for Tactile Internet (CeTI)}
 \institution{Technical University Dresden}
 \streetaddress{Nöthnitzer Str. 46}
 \city{Dresden}
 \state{}
 \country{Germany}
 \postcode{01187}
}
\email{simon.hanisch@tu-dresden.de}

\author{Thorsten Strufe}
\affiliation{%
 \department{KASTEL Security Research Labs}
 \institution{Karlsruhe Institute of Technology}
 \streetaddress{Am Fasanengarten 5}
 \city{Karlsruhe}
 \state{}
 \country{Germany}
 \postcode{76131}
}
\email{thorsten.strufe@kit.edu}

\begin{abstract}
Face images are a rich source of information that can be used to identify individuals and infer private information about them. 
To mitigate this privacy risk, anonymizations employ transformations on clear images to obfuscate sensitive information, all while retaining some utility. 
Albeit published with impressive claims, they sometimes are not evaluated with convincing methodology.

Reversing anonymized images to resemble their real input --- and even be identified by face recognition approaches --- represents the strongest indicator for flawed anonymization.
Some recent results indeed indicate that this is possible for some approaches. 
It is, however, not well understood, which approaches are reversible, and why.
In this paper, we provide an exhaustive investigation in the phenomenon of face anonymization reversibility.
Among other things, we find that 11 out of 15 tested face anonymizations are at least partially reversible and highlight how both reconstruction and inversion are the underlying processes that make reversal possible.
\end{abstract}

\keywords{anonymization, evaluation methodology, reversibility}

\maketitle

\section{Introduction}

In today's world, biometric data is pervasively captured as more sensors are recording us in larger quantity and quality. 
Take, for example, the increasing usage of surveillance cameras, autonomous vehicles that scan their surroundings, mixed reality devices, or sensors in various smart devices. 
This development poses challenges to individual privacy, as extensive sensitive information can be inferred from our biometric data.
Examples are abound, and they include identity \cite{deng_arcface_2019, Wan_Gait_Recognition_2018}, personal preferences \cite{kroger2020does}, sexuality \cite{rieger2012eyes, johnson2007swagger}, health status \cite{LIU2021107187} and medical conditions \cite{kroger2020does}. 
Some suggested biometric data protection techniques attempt to prevent this threat.
One class of these systems aims at irreversibly transforming face images in such a way that privacy-sensitive inferences are no longer possible, while trying to retain the utility of the image. 
There are many proposals \cite{meden2021privacy} on how to design such anonymizations, however, the evaluation methodology to quantify how much privacy protection they offer is still lacking.

We observe a severe problem with the common evaluation methodology: they frequently rely on weak attacker models that assume an attacker is unaware of the anonymization.
An attacker who is aware of the modifications is stronger (and we claim: more realistic!) as they can actively try to remove the protection.
Today, the most common method to build such an attacker is to train recognition systems on protected data (e.g.~\cite{mcpherson_defeating_2016}).
This helps the recognition system to adapt to changes caused by the anonymization.

However, we argue that training recognition systems on protected data is not optimal, as these systems were never designed to work on protected data. Instead, we pursue an alternative direction in which the anonymized image is attempted to be reversed to its clear image in an intermediate step before the recognition system performs the identification. Preventing reversal is a key requirement for biometric privacy, but it is often overlooked in evaluation.
For an anonymization to protect individuals, it must be a one-way-function for any arbitrary adversary and therefore reversibility is the worst-case failure of such a protection technique.

The literature for face images \cite{tekli_framework_2019, ruchaud_automatic_2016} already has shown that specific reversing techniques such as deblurring, denoising and super-resolution can be successful at reversing basic anonymizations.
A recent paper by Hao et al. \cite{hao2020robustness} attempts to use a general purpose machine learning model for a variety of face anonymizations and achieves higher identification accuracies, showing that some of them are reversible.
These initial results show that there might be a general approach to reversing anonymizations. We are the first to investigate in-depth if reversibility is a widespread problem and try to assess what makes some anonymizations reversible.
It is still an open research question which (groups of) anonymizations are reversible, to what extent reversibility generalizes, and how the evaluation of identification on reversed data compares to the common evaluation methodology.

Our main contribution in this paper is an exhaustive investigation of the phenomenon of face anonymization reversibility.
We try to answer the question: How and when can face anonymization techniques be reversed?
For this, we design and conduct a large number of experiments that investigate different aspects of this question.
In particular, we consider the following:
\begin{itemize}
    \item We define an evaluation methodology that uses a general de-anonymization before face recognition and test it on a large number of face anonymizations. This allows us to systematically investigate which (groups of) anonymizations are reversible;
    \item To investigate what makes reversal possible we design and test a general machine learning model based on two underlying processes: reconstruction and inversion;
    \item We test specialized and general de-anony\-mizations, as well as the common anonymization evaluation methodology on a large number of face anonymizations to highlight differences between them;
    \item We test cases where training and test data does not match or is from different data sets to investigate to which extent de-anonymizations generalize;
    \item  We conduct a user study to assess the perceived visual appeal of anonymized images to investigate if there is a trade-off between reversibility and utility.
\end{itemize}

\section{Background}\label{sec:background}

Here we present the background and terminology which is required to understand our work and the assumptions it is based on. 

Following established vocabulary~\cite{ISO/IECJTCSC37} we use biometric characteristics to describe the biological and behavioral characteristics that can be used to extract biometric features which in turn can be used by \textbf{biometric recognition} to identify individuals or infer attributes, such as age~\cite{DEHSHIBI20102431} and sex~\cite{pollick_gender_2005} about them.
To prevent biometric recognition \textbf{privacy enhancing technologies (PETs)} are employed which obfuscate the private information in the data from internal and external observers. The specific term of \textbf{anonymization} refers to PETs which remove all identifiers that directly identify individuals. Anonymization takes biometric \textbf{clear data} as input and outputs \textbf{anonymized data}. While the use cases for anonymization can vary widely, by definition their output is always anonymous, i.e. it is impossible to identify individuals from the data. For the remainder of this work, we will assume a data publishing model as our system model, as such the biometric data must be anonymized before it is published to a third party. This implies that the anonymization must not be reversible as else a malicious third party can simply remove the anonymization to access the data. An example of this system model would be a user who anonymizes their data before uploading it to a social media site, for example anonymizing their own faces or the ones of bystanders. We will call anonymized data on which anonymization reversal was attempted \textbf{de-anonymized} data.

\subsection{Anonymization Evaluation State of the Art}
The most common evaluation methodology today to test the anonymization of biometric data is to measure the recognition accuracy with a biometric recognition system. By comparing the accuracy of the clear and anonymized data the protection of the anonymization can be determined.

Newton et al.~\cite{newton_preserving_2005} proposed to differentiate these experiments by which data was used for enrollment and testing of the recognition system.
In their approach, \textbf{naive recognition} uses clear data as enrollment data, and then anonymized data is used as test data.
\textbf{Parrot recognition} on the other hand enrolls the recognition system on anonymized data before it is tested against anonymized data, which most of the time improves performance as the recognition system can adapt to the anonymized data better. 
The parrot recognition approach was further improved by Srivastava et al.~\cite{srivastava_evaluating_2020} who split the parrot recognition case into a semi-informed attacker who only knows the anonymization method but not its parameters and an informed attacker who knows the anonymization method and its exact parameters.
In addition, McPherson et al. \cite{mcpherson_defeating_2016} adapt the recognition model to three anonymizations by training it on the specific anonymized data.
A recent initiative to build a common methodology how to evaluate speaker anonymization is the VoicePrivacy~\cite{tomashenko_introducing_2020} challenge. 
Similar to the methodologies above, they define the attackers by how much access to anonymized training data they have.

A different evaluation approach considers the reversibility of anonymizations.
Early versions \cite{tekli_framework_2019, ruchaud_automatic_2016} used de-anonymizations specific for individual anonymizations to test their reversibility.
More recently, Hao et al. \cite{hao2020robustness} used the general image-to-image machine learning model Pix2Pix \cite{isola2017image} to attack multiple anonymizations. While their initial results indicate that general reversal might be possible, they are not well understood and limited in the number of anonymizations considered.
Besides evaluating against a biometric recognition system it is also possible to evaluate against human evaluators who attempt to recognize individuals, as done in \cite{lander_evaluating_2001}.
However, this is less common because it is much easier to run automated biometric recognition methods than to conduct user studies.
Also, McPherson et al. \cite{mcpherson_defeating_2016} and Hao et al. \cite{hao2020robustness} claim that humans may no longer be the gold standard for human identification.

\section{Related Work}\label{sec:related_work}
\textit{Template protection} is closely related to biometric data anonymization as its goal is to remove all attributes, except the identity, from the data. ISO-24745~\cite{ISO24745} requires template protection schemes to be irreversible. Cappelli et al.~\cite{cappelli2007fingerprint} reconstructed fingerprints from templates. De-anonymization attacks are a common threat to biometric template schemes as a survey by Gomez et al.~\cite{gomez2020reversing} shows. Biometric data anonymization schemes share the same system model as template protection schemes and hence also must be irreversible to protect user privacy. However, the attacks on template protection schemes are not directly applicable as template protection schemes try to keep the identity of a subject while anonymizations try to remove it. 

\textit{Evaluation methodology improvement} is a common research subject, that is not only explored for biometric data anonymization but also in the field of biometric recognition.
Philips et al.~\cite{phillips2012good} suggested partitioning of the used biometric data set according to the quality of the data samples.
The reasoning for this methodology is that it becomes easier to judge the robustness of recognition algorithms.
Stolerman et al.~\cite{stolerman2013classify} looked critically at the usage of a closed-world assumption for stylometry recognition.
They found that many stylometry methods fail when an open-world assumption is utilized.
Goga et al.~\cite{goga2015reliability} were able to show that the matching of profiles across social networks is not as easy as previously thought by making the assumptions in their evaluation more realistic.
Arp et al.~\cite{arp2022and} had a look at the used methodologies for using machine learning in the security field and identified common mistakes.
Hanisch et al.~\cite{hanisch2023false} investigated how the specific selection of the identities of the evaluation data set can be used to create a more challenging evaluation data set for biometric anonymization.
Wenger et al.~\cite{Wenger_2023_Anti-Facial} performed a systematization of knowledge of face anonymization techniques that focus on preventing online face recognition. As one of the design properties of face anonymization, they identify the longer-term robustness of the technique, thus taking into account that face recognition techniques evolve and get better over time.
All these works highlight that it is important to critically look at the used evaluation methodologies to further drive the development of the field e.g. anonymizations towards better privacy protection.

\textit{Specialized reversibility attacks} for biometric data anonymization techniques have been proposed in the past. Xu et al.~\cite{shen_deep_2018} train a convolutional neural network to reconstruct blurred faces. Lu et al.~\cite{ma_deep_2020} have proposed a super-resolution approach that removes Pixelation from face images. A denoising and deblurring approach was proposed by Zamir et al.~\cite{zamir_multi-stage_2021} who use an auto-encoder to recover a restored version of an image. Further methods performing deblurring are by Krishnan et al.~\cite{krishnan_blind_2011}, Pan et al.~\cite{pan_deblurring_2014}, and Tsai et al.~\cite{tsai_stripformer_2022}. Tekli et al.~\cite{tekli_framework_2019} have created a framework that evaluates image anonymization and can apply three different specialized reversibility attacks on the images. While for this specific use-case of deblurring and denoising images methods exist it is not clear how they compare to a general reversibility attacker. Missing is also a systematic evaluation of how the reversibility approach works against various types of anonymization. 
\section{Evaluation Methodology}\label{sec:approach}
In this section, we first explain why we require the evaluation of reversibility and then define our attacker model.
We will use the resulting evaluation methodology throughout this paper.

\subsection{Analysis} 
As we already mentioned in the introduction, most evaluations of biometric data anonymizations assume a weak attacker which is not aware of the anonymization that was performed on the data. This is an unrealistic limitation of the attacker as anonymizations are often easy to detect (e.g. a blurred face) and we assume that a dedicated attacker will always be able to detect that the data is anonymized. Further, for PETs, we are most commonly interested in worst-case performance, so assuming a strong attacker is only natural.

 A strategy~\cite{tomashenko_voiceprivacy_2022, newton_preserving_2005} that has been proven to be successful for  worst-case evaluation is the retraining of biometric recognition systems using anonymized samples to adapt the model to the anonymization. However, biometric recognition systems have never been designed for dealing with anonymization and hence we expect that a dedicated approach to reverse the data anonymization can be more successful. Looking at the literature \cite{tekli_framework_2019, ruchaud_automatic_2016} we find that some specialized approaches to reverse anonymizations already exist (e.g. deblurring) and are successful, indicating that de-anonymization attacks might be possible against other anonymizations. However, developing specialized approaches for every anonymization would be time-consuming and the results would not be directly comparable across anonymizations.
Recently, Hao et al. \cite{hao2020robustness} used Pix2Pix \cite{isola2017image} as a general de-anonymization to test reversibility of some anonymizations. Pix2Pix is a general image-to-image translation model that can be trained with any set of image pairs. By training it using pairs of anonymized and clear images, the reversal can be learned and later applied to anonymized images. However, their experiments are limited in the anonymizations considered and leave open how and when exactly reversal is possible and how it compares to naive and parrot recognition. In this work, we want to investigate these questions in detail and therefore define our attacker model and the resulting reversibility evaluation methodology in the following.

\subsection{Attacker Model}
The \textbf{goal} of our attacker is to identify individuals in anonymized biometric recordings, by reversing the anonymization of the data.
To achieve this goal, the attacker \textbf{knows} that the recordings are anonymized, however, in order to make the attacker general and agnostic to the anonymization we regard the anonymization as a black box for which neither the parameters nor the anonymization method itself are known.
This information is additionally known in the case of specialized de-anonymizations.
Further, the attacker has access to a \textbf{clear data} set of biometric recordings which does not have to include the individuals under attack and therefore could for example be a large publicly-available research data set.
This set can be anonymized using the black box anonymization (similar to encryption oracles in cryptography) which results in a \textbf{corresponding anonymized data set}. This is based on the assumption that generally an attacker can detect and identify the anonymization used and apply it themselves.
For the identification of individuals the attacker also possesses clear \textbf{enrollment data} and anonymized \textbf{test data}, as in the common methodology. Since our adversary should evaluate the robustness of the anonymization against being reversed, we assume a pessimistic scenario for the anonymization. This means that the anonymization must work even in a worst-case scenario. This in turn means that we pick an easy identification scenario as this is a hard anonymization scenario~\cite{hanisch2023false}. For our attacker, this means that the face images to be anonymized and then attacked are of high quality and contain clearly identifiable faces~\cite{reshma_approaches_2019}. We assume that if the anonymization is not reversible on these high quality images, it will not be reversible on lower quality images.

\begin{figure}[h!]
    \includegraphics[width=\linewidth]{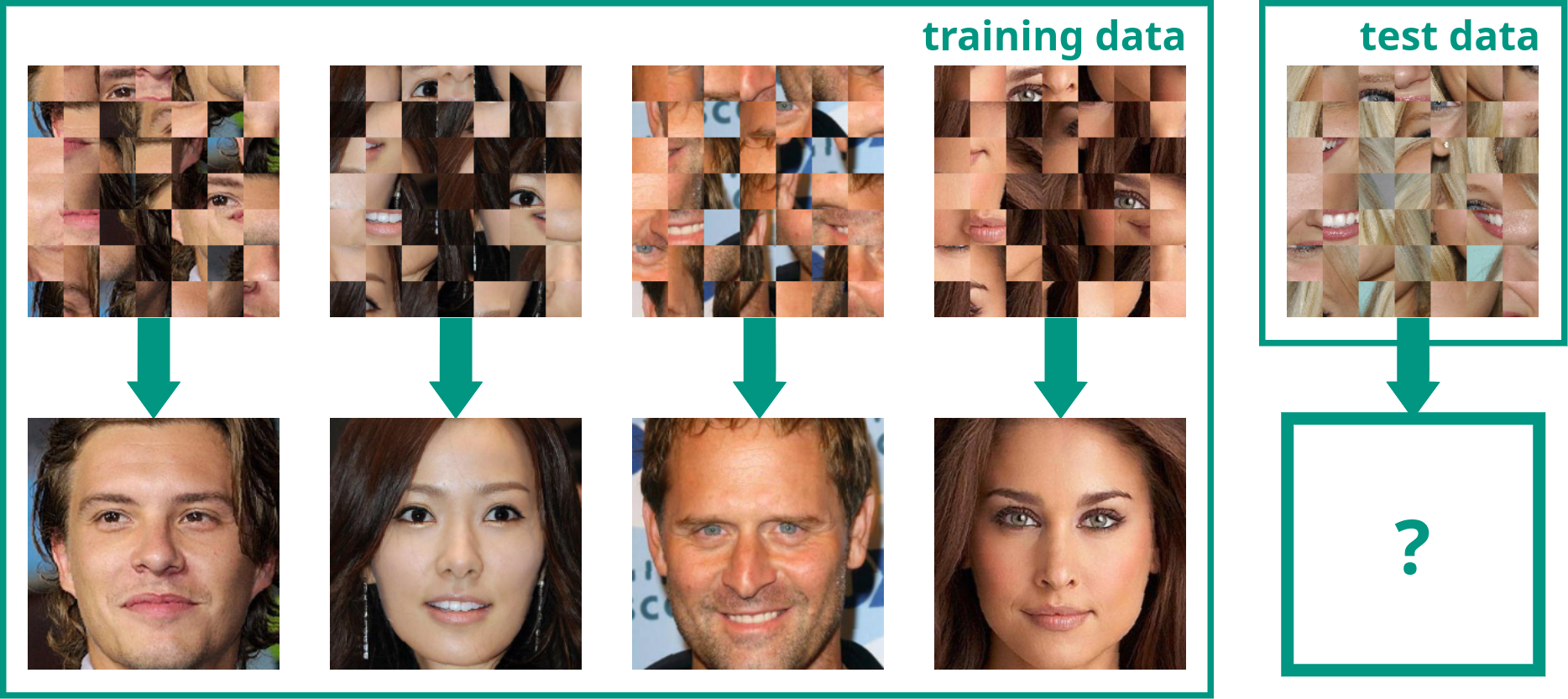}
    \caption{Data access of the attacker model. For training, the model has access to both anonymized and respective clear images, for testing only anonymized images are available.}
    \label{fig:de-data}
\end{figure}

A visual example of the data sets in our attacker model can be found in Figure~\ref{fig:de-data}. The success of the attack will be measured by how well the attacker can identify the individuals in the test data set (not how well the anonymized data was de-anonymized). We consider the attacker to be successful if they can identify individuals in the de-anonymized recordings more successfully than in the anonymized recordings. 
A comparison of our attacker model to existing ones can be found in Table~\ref{tab:cmp-attackermodels}.

A simple real-world example of our attacker would be an attacker that tries to identify individuals on a social media site that anonymizes faces in images (e.g. the faces of bystanders in the background) before they are shared online. By uploading its clear data set, the attacker receives the corresponding anonymized data set and can then perform its de-anonymization attack.

\begin{table}[h!]
\centering
\caption{Comparison of attacker models in regards to which information and data they have access to.}
\small\setlength{\tabcolsep}{2.5pt}
\begin{tabular}{l|cccc}
  Model & naive & parrot & special. & \textbf{ours}\\
  \hline
  Knowledge of ... \\
  ... manipulation & \xmark & \cmark & \cmark & \cmark \\
  ... manipulation method & \xmark & \xmark & \cmark & \xmark \\
  ... manipulation parameters &  \xmark & \xmark & \cmark & \xmark \\
  Access to data pairs & \xmark & \xmark & \xmark & \cmark\\
\end{tabular}
\label{tab:cmp-attackermodels}
\end{table}

\subsection{Reversibility Evaluation}
Based on our attacker model, we design an evaluation methodology to test reversibility of biometric anonymization. The idea of the methodology is to perform general de-anonymization before the identification is tested on the data. To keep the de-anonymization general we keep it agnostic to the anonymization under test by using machine learning to learn a model that transforms the anonymized data back into its corresponding clear data and therefore de-anonymizes the data. This way the attacker can be easily adapted to any anonymization, simply by the training data of the de-anonymization being anonymized using the specific anonymization method that is being evaluated.

\begin{figure}[h!]
    \includegraphics[width=\linewidth]{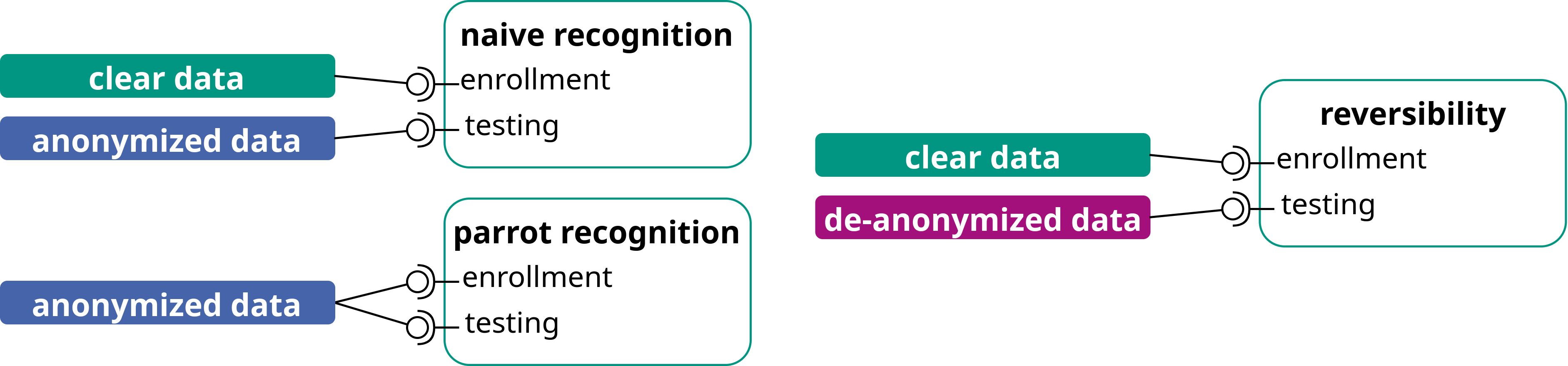}
    \caption{Recognition attacker models and their respective data usage for training and testing the biometric recognition system they use for their attack.}
    \label{fig:de-recognition}
\end{figure}

After the training of the model, we use it to de-anonymize the test data. To now perform the identification we use a biometric recognition system in which we enroll clear data samples of the individuals we wish to identify and test against the de-anonymized test data (for a comparison to previous methodologies, see Figure~\ref{fig:de-recognition}). We select clear data as the enrollment data because due to the de-anonymization the data is closer to clear than anonymized data. This assumption was confirmed in an experiment, in which the average accuracy (Facenet, VGG-Face2 \& ArcFace) for all fifteen anonymizations was 49.2\% with clear data and 27.1\% with anonymized data for enrollment with data de-anonymized using our approach (see Section~\ref{sec:design}) as test data. The identification accuracy of the recognition system on the de-anonymized data is a metric of the anonymization's ability to protect the privacy of individuals in the biometric recordings. If the recognition system is able to identify individuals, then either anonymized data is sufficient to identify individuals (the case caught by previous evaluation methodology) or the anonymization was reversible.

\section{Design}\label{sec:design}
For our investigation into the phenomenon of face anonymization reversibility, we want to better understand what makes reversal possible. To do this, we design a new machine learning model that is specifically designed for general de-anonymization and not based on previous models like Pix2Pix~\cite{isola2017image} which have not been originally been designed to reverse anonymizations. Hence, Pix2Pix can only demonstrate that reversibility is possible but does not allow us to reason why reversing anonymizations is possible. We do not necessarily want to create the best-performing model, but rather purposely design a model that helps us understand the phenomenon of reversibility.

For the design, we are guided by two underlying processes: reconstruction and inversion. Reconstruction exploits the correlations and dependencies in the biometric data to recover removed information. Take for example face images in which due to the structure of the face it is clear where the position of the eyes is, or how the color of one eye most of the time also gives you the color of the other eye. Inversion on the other hand is the direct undoing of the operation that the anonymization performed on the data. While reconstruction will always result in small differences to the original (lossy), inversion can also perfectly reverse (lossless).
A model trained to de-anonymize anonymized data will use a combination of both.

Considering that both our input and output are images, we decide to select an under-complete auto-encoder as the base model. Auto-encoders compress the input into a small latent code that represents the input before decoding it back into the same domain as the input making them popular choices as a method to remove noise from images called denoising auto-encoders~\cite{gondara_medical_2016, song_image_2020}.
The benefit of auto-encoders is that the encoder and decoder learn the intrinsic dependencies in the data which can help with the reconstruction of data that was obfuscated by anonymization. A specialized version of auto-encoders that use this ability are auto-encoders which are used as generators for deepfakes~\cite{nguyen_deep_2022, mirsky_creation_2022}.

For denoising, we find both auto-encoders with linear and convolutional layers being used. Many common face anonymizations perform localized changes in the image and therefore convolutional layers with their locality and translation invariance properties seem like the obvious choice. In these cases, the dominant process is reconstruction. Convolutional layers are also the more common option whenever dealing with images, since there is the concept of neighborhoods and relative positions of pixels as opposed to linear layers that rather work with vectors and interpret them as simple lists of values. In situations in which convolutional layers can solve a problem, they should also generally be preferred over linear layers as they have fewer trainable parameters which will speed up the training process.

Our attacker's machine-learning model is supposed to be general. In other words, it should be able to reverse any anonymization. 
While many anonymizations perform only local modifications, some apply global changes to the image, as for instance permutations.
It hence is not sufficient to use convolutional layers, with local effects, but functionality to invert global changes has to be implemented.

Such a scenario is not considered by the convolutional-only architecture of Pix2Pix \cite{isola2017image}, the model used for reversal by Hao et al. \cite{hao2020robustness}. %

\begin{figure}[h!]
    \centering
    \includegraphics[width=.8\linewidth]{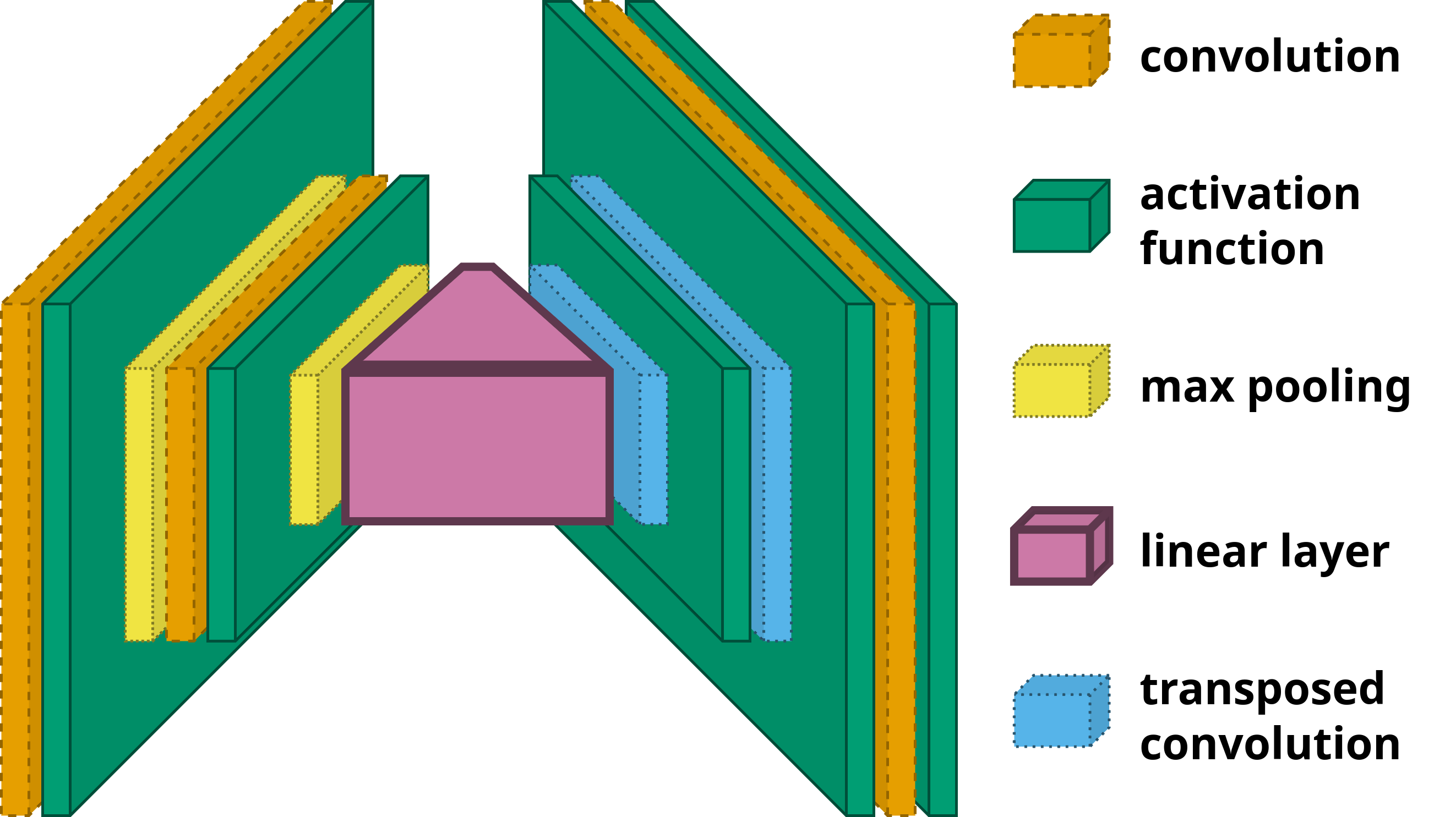}
    \caption{Design of our machine learning model}
    \label{fig:de-model}
\end{figure}

In linear layers, the locality principle does not exist and outputs can depend on any (or all) inputs including those that would not be considered close enough by a convolutional layer. Therefore, a machine learning model that is actually general, would use linear and not convolutional layers. However, linear layers require memory proportional to input size times output size. Considering that we are working with high-resolution RGB images we choose to use a model with a single linear layer between the encoder and decoder to keep the model size feasible. A visual representation of the described model is shown in Figure~\ref{fig:de-model}.

In the encoder part, the model uses two convolutional layers with following activation functions and max pooling layers. The max pooling layers reduce the dimension of the input, each of them halving the width and height of the image. The decoder is designed symmetrically: two transposed convolutional layers followed by activation functions.
Each of them quadruples the number of pixels, resulting in an output resolution that matches the input.

As we are using RGB images, our input data has three channels.
The first convolutional layer of the encoder increases this to a specified number of features. We consider this number of features a hyperparameter for which we conduct experiments to find a suitable value. However since the number of features influences the size of the linear layer, it is limited by the available GPU memory. To reduce the number of channels back to three in the output, the decoder part also includes a convolutional layer after the two transposed convolutional layers. For the activation function we considered Sigmoid, Tanh, and ReLU (rectified linear unit), but empirically found LeakyReLU to perform best.

Similarly, we also test multiple options for loss functions to be used during model training. This includes standard regression loss functions such as mean squared error (MSE) and mean absolute error (MAE) as well as computer vision-specific ones like structural similarity (SSIM)~\cite{wang_image_2004}. We acknowledge that more advanced loss functions such as an identity loss function that reduces the difference in recognized identity rather than the difference in pixel values might also be very suitable in this use-case, but choose to keep this general de-anonymization purposely simple to be able to understand the results better.

\section{Techniques}\label{sec:techniques}
In this section, we introduce all the anonymization and de-anonym\-ization techniques that we use in our experiments.
For each, we consider both commonly used basic methods as well as state-of-the-art approaches.
We make sure that our selection of methods covers all categories that are relevant for our scenario.

\subsection{Anonymizations}
\begin{figure*}[ht!]\centering\setlength{\tabcolsep}{2pt}
\begin{tabular}{*{8}{>{\centering\arraybackslash}m{2.05cm}}}
  \includegraphics[width=2cm]{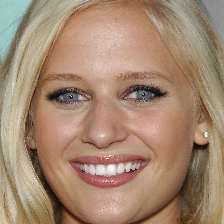} & \includegraphics[width=2cm]{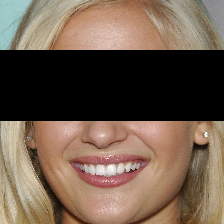} & \includegraphics[width=2cm]{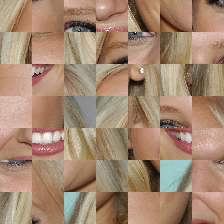} & \includegraphics[width=2cm]{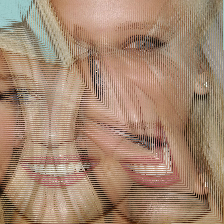} & \includegraphics[width=2cm]{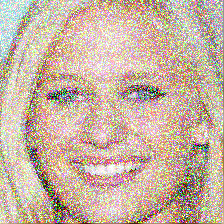} & \includegraphics[width=2cm]{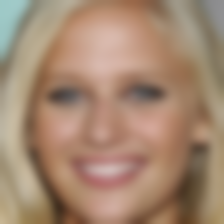} & \includegraphics[width=2cm]{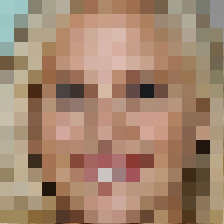} & \includegraphics[width=2cm]{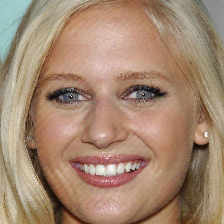} \\
  Clear & Eye Mask & Block Permut. & Pixel Reloc. & Gauss. Noise & Gauss. Blur & Pixelation & Fawkes\\
  \\
  \includegraphics[width=2cm]{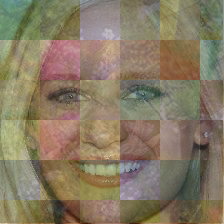} & \includegraphics[width=2cm]{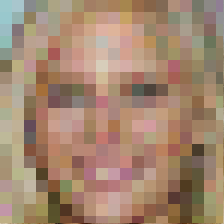} & \includegraphics[width=2cm]{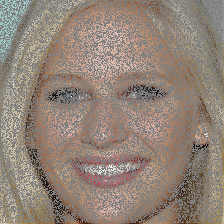} &  \includegraphics[width=2cm]{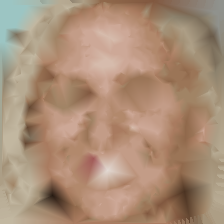} & \includegraphics[width=2cm]{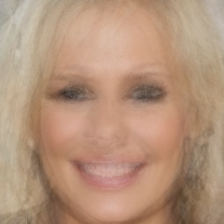} & \includegraphics[width=2cm]{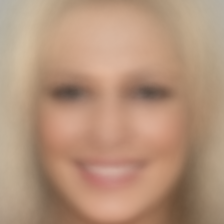} & \includegraphics[width=2cm]{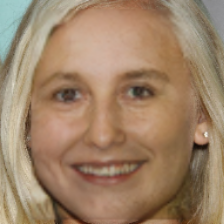} & \includegraphics[width=2cm]{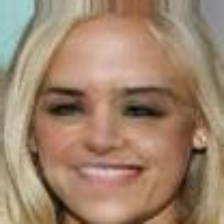}\\
  k-RTIO & DP Pix & DP Snow & DP Samp & \textit{k}-Same-Pixel & \textit{k}-Same-Eigen & DeepPrivacy & CIAGAN\\
\end{tabular}
\caption{Different face anonymization methods we consider}
\label{fig:anonymization}
\end{figure*}

For all introduced anonymizations, an example image can be found in Figure~\ref{fig:anonymization}.

\subsubsection{Basics}
Basic anonymizations are the most commonly used methods as they are easy to implement and often provide straightforward parameters to control the privacy-utility trade-off. Their main utility goal is to keep the image similar to the original one.

\textit{Eye Mask.} The pixels in the eye area of the face are removed and replaced by a black bar.

\textit{Block Permutation.} The face image is split into equally-sized blocks which are then permuted.
The same permutation is used for all images. Note that we add Block Permutation as a trivial example of reversible anonymization in order to test our de-anonymization methodology.

\textit{Pixel Relocation \cite{cichowski_reversible_2011}.} Cichowski and Czyzewski introduce an anonymization designed for videos that is based on relocating individual pixels using a fixed permutation. It is designed to be reversible when a secret key is known.

\textit{Gaussian Noise.} For every pixel of every channel in the image, random noise is drawn from a Gaussian distribution and added to the pixel's value.

\textit{Gaussian Blur.} The face area of the image is blurred using Gaussian blur. This is done by performing a convolution on the image with a Gaussian kernel matrix.

\textit{Pixelation.} The resolution of the image is reduced. The parameter is the number of remaining pixels on either axis.

\subsubsection{Adversarial Machine Learning}
Anonymizations in this category achieve their privacy protection by attacking the face recognition machine learning models that are used to identify individuals.
These data poising attacks have been criticized as they target specific face recognition models and therefore do not offer any protection anymore when new models get implemented in the future~\cite{radiya-dixit_data_2021}.
As these methods explicitly do not protect against humans, the term 'anonymization' may be considered incorrect (rather: de-identification or anti-facial-recognition).
Nevertheless, due to their similarity, we add one such method to our comparison: \textit{Fawkes~\cite{shan_fawkes_2020}} adds "imperceptible pixel-level changes" to face images.
Fawkes’ use-case assumes that the anonymized images are used to train the recognition system and can therefore "poison" the information base so that later recognition attempts on non-anonymized data fail. Its utility goal is to allow human observers to still recognize the person in the image.
The idea is to compute minimal perturbations for an image that cause significant changes in the output of the face recognition model.
We use the open-source implementation by Fawkes' authors Shan et al.

\subsubsection{Overlay}
\textit{k-RTIO~\cite{rajabi_impracticality_2021}} (K-randomized transparent image overlays) adds a semi-transparent overlay to the face image.
Based on the image’s identifier and a secret key, images from a known overlay image data set are selected.
The overlay images are then block permuted based on the secret key and combined.
This combination is overlayed on the face image.
This anonymization is designed to be reversible with the knowledge of the secret key.
The use case is the disruption of face recognition systems that may run on cloud hosted images while preserving enough utility so that the anonymized images might still be usable in the cloud environment without the need to download and de-anonymize them.

\subsubsection{Differential Privacy}
A commonly used framework in anonymizations is Differential Privacy (DP) which allows formal and provable privacy guarantees.
By definition, an adversary cannot effectively distinguish between the outputs of a differentially private mechanism.
In the case of face anonymization, this would theoretically guarantee that images cannot be re-identified by a face recognition method. The utility goal in this category generally is to keep the image similar to the original one.

\textit{DP Pix~\cite{fan_image_2018}.} The image is first pixelated by averaging the pixels within blocks.
Then a Laplace perturbation is added to the pixelized image.
The algorithm was originally designed for grayscale images, we adapt it to RGB images by performing all algorithm steps for each channel separately.
We implement DP Pix ourselves using the description in \cite{fan_image_2018} and the pseudo-code in~\cite{reilly_comparative_2021}.

\textit{DP Snow~\cite{john_let_2020}.} A configurable percentage of pixels in the image are replaced with gray pixels.
When $\delta$ is the percentage of replaced pixels, this anonymization is $(0, \delta)$-differentially private according to John et al.~\cite{john_let_2020}. The stated utility goal is to preserve the landmarks of the face.

\textit{DP Samp~\cite{reilly_comparative_2021, wang_videodp_2019}.} This method was originally proposed for video anonymization in~\cite{wang_videodp_2019} and adapted for grayscale images in~\cite{reilly_comparative_2021}.
We further adapt it for RGB images.
Our variant works as follows: K-Means is used to generate $k$ clusters from the pixels of the image.
For each cluster, the number of pixels within a threshold to the mean cluster color is counted.
Based on these frequencies, each cluster is allocated a fraction of the overall privacy budget.
The privacy budget of a cluster determines how many pixels within the cluster are randomly sampled.
The sampled pixels from all clusters are then used to linearly interpolate the remaining pixels for the final anonymized image.
We implement DP Samp ourselves based on the pseudo-code in~\cite{reilly_comparative_2021}.

\subsubsection{\textit{k}-Anonymity}
A different formal framework that allows for privacy guarantees is \textit{k}-anonymity.
Its basic idea is to modify the data in such a way that any single point is equally likely to belong to any of \textit{k} identities.
This reduces the re-identification accuracy to a theoretical maximum of $1/k$.
Utility is achieved by grouping individuals that share similar attributes, resulting in the anonymized image preserving these attributes.

For face anonymization, \textit{k}-anonymity was first formalized and proposed by Newton et al. in~\cite{newton_preserving_2005}.
They however make some assumptions that are unsuitable for our use-case (and many real-world scenarios) including that there is only a single image per identity and no other images or identities are added after the initial anonymization \cite{meden_-same-net_2017, muraki_anonymizing_2013}.
Further, there is no straightforward way to split the anonymized data set into multiple parts without breaking the formal privacy guarantee.

We therefore implement a variation on their approach.
We create an anonymization background data set that contains the images of identities which are not used anywhere else in our framework.
We train a PCA on the images in this data set and save their representations in a database.
When anonymizing an image, we find the $k-1$ closest images in the PCA-space (only one per identity).
The anonymized image is then the average of the original image and the $k-1$ closest images from the background data set.
This allows us to anonymize multiple images for the same person and to later split the anonymized data set into multiple parts without adverse effects.
For \textbf{\textit{k}-Same-Pixel~\cite{newton_preserving_2005}} the $k$ images are averaged in the pixel-space while for \textbf{\textit{k}-Same-Eigen~\cite{newton_preserving_2005}}, the anonymized image is the inverse transform of the average of the $k$ images' PCA representations.

\subsubsection{Synthesis}
Anonymizations in this category replace the entire face in the image with a new synthetic one.
Since this removes the majority of identifying features of the original face, recognition systems fail to match these images to the correct person.
At the same time, utility can be achieved by creating synthetic faces that preserve specific attributes of the original one.

\textit{DeepPrivacy~\cite{hukkelas_deepprivacy_2019}.} Hukkelås et al. use a conditional generative adversarial network which considers original pose and background of the image. It has the goal to preserve a variety of attributes of the original face while protecting the privacy of the individual. We use the authors' open-source implementation.

\textit{CIAGAN~\cite{maximov_ciagan_2020}.} The approach by Maximov et al., CIAGAN, is based on conditional generative adversarial networks together with a novel identity control discriminator. The goal is to remove identification characteristics of people while keeping the necessary features required for detection, recognition and tracking. Anonymized images are supposed to be high-quality and realistic for human observers. We use the authors' open-source implementation.

Additional examples in this category include AnonFaces \cite{le_anonfaces_2020} and StyleID \cite{le_styleid_2023} which we did not include as we expect them to show the same results as DeepPrivacy and CIAGAN.

\subsection{De-Anonymizations}
In the following, we want to introduce de-anonymizations which can be compared to our general de-anonymization model.
\subsubsection{Basics}
Basic de-anonymizations are tools from the area of image processing and have not been specifically designed to reverse biometric anonymizations.
However, they can still improve recognition results for a wide range of anonymizations.

\textit{Linear/Bicubic interpolation.} For Pixelation, we simply use linear or bicubic interpolation to upsample the image back to its original size. For any other anonymization, the images are first downsampled and then back up using linear or bicubic interpolation. The intermediate resolution is determined by calculating the SSIM of the re-upsampled image and the original clear image for all images in the de-anonymization training data set for a variety of intermediate resolutions. The intermediate resolution that achieves the highest average SSIM is then used on the test data set.

\textit{Wiener filter~\cite{hunt_matrix_1971}.} This applies a wiener filter to the image. We test both a version where the parameters are determined by testing them on the training data set and choosing the ones with highest SSIM and version based on~\cite{orieux_bayesian_2010} that blindly estimates the parameters for every image individually. We use the implementations from the scikit-image library~\cite{van_der_walt_scikit-image_2014}.

\textit{Richardson-Lucy Deconvolution~\cite{richardson_bayesian-based_1972, fish_blind_1995}.} This applies a Richardson-Lucy deconvolution on the image. Parameters are blindly estimated on a picture by picture basis using the approach from~\cite{fish_blind_1995} and the implementation from the scikit-image library~\cite{van_der_walt_scikit-image_2014}.

\textit{Wavelet denoising~\cite{chang_adaptive_2000}.} This applies the adaptive wavelet thresholding for image denoising approach by Chang et al. as implemented in the scikit-image library~\cite{van_der_walt_scikit-image_2014}.

\subsubsection{State-of-the-art Approaches}
Anonymizations that blur, pixelate or add noise to images are very similar to processes that naturally happen to photos that reduce their quality.
Significant amounts of research have been done to mitigate these natural processes which have resulted in deblurring, super resolution and denoising approaches.
We can use state-of-the-art approaches from these areas as de-anonymizations for our artificially degraded images to improve recognition accuracy.
We planned to use face specific approaches such as~\cite{shen_deep_2018} and~\cite{pan_deblurring_2014}, however we were unable to get the authors' open-source implementations working.

\textit{Deep-Face Super-Resolution~\cite{ma_deep_2020}.} This face super resolution approach uses two recurrent neural networks with iterative collaboration for face image recovery and landmark estimation. The goal is to recover high-quality face images from low-resolution images. We use the authors' open-source implementation and abbreviate it as "DIC SR".

\textit{Blind deconv. using a normalized sparsity measure approach~\cite{krishnan_blind_2011}.} Here, a mathematical model is used to reverse blurring on images without knowledge of the used blurring method. We use the authors' open-source implementation. We abbreviate this method as "Norm sparsity".

\textit{MPRNet~\cite{zamir_multi-stage_2021}.} Using a machine learning model with a multi-stage architecture using encoder-decoder pairs in combination with a high-resolution branch that retains local information, MPRNet attempts to restore high-quality images from degraded inputs. The authors provide pre-trained models for denoising and deblurring as well as an open-source implementation that we use.

\textit{Stripformer~\cite{tsai_stripformer_2022}.} Blurred images are restored using a machine learning model with a transformer-based architecture. We use the authors' open-source implementation as well as the model which they trained on the GoPro dynamic scene deblurring data set by Nah et al.~\cite{nah_deep_2017}.

\textit{Pix2Pix \cite{isola2017image}.} Using a conditional neural network, deep learning is used for general image-to-image translation. It is used by Hao et al. in \cite{hao2020robustness} to test the reconstruction of obscured face images.

\subsubsection{Specialized Approaches}
For some anonymizations, specialized approaches for the exact anonymization that was used can be implemented.

\textit{Interpolation.} For the anonymization DP Snow, we interpolate every completely gray pixel from its eight neighboring pixels while ignoring any neighboring completely gray pixels. Considering the high-resolution property of the used images, this makes the reasonable assumption that neighboring pixels have similar colors and that hard edges are rare in natural face photos.

\textit{Learn permutation.} For Block Permutation and Pixel Relocation, we can use the access to training images with the exact same permutation to learn this permutation and then apply the inverse on the test images. This works by matching the pixel colors from clear to anonymized images.

\section{Experiments}\label{sec:evaluation}

In this section, we design, conduct and show the results of our experiments that evaluate reversibility.
We first present the expectations that we want to test based on the aspects of reversibility that we want to investigate. Then we explain the corresponding experiments and their results.
The evaluation of utility can be found in Section~\ref{sec:userstudy-utility} and finally the comparison with human observers can be found in Appendix~\ref{sec:userstudy}.

\subsection{Expectations}
Our main goal is an exhaustive investigation in the phenomenon of anonymization reversibility.
For this, we consider the main aspects that we presented in the introduction.
For each, we determine expectations which we then test in experiments.

The first aspect considers which anonymizations or groups of anonymizations can or cannot be reversed.
\textbf{E1.1:} We expect that permutations (Block Permutation, Pixel Relocation) can be perfectly reversed, meaning that we recover the exact pixel by pixel clear image. This is because these anonymizations do not actually remove any information from the image.
\textbf{E1.2:} As synthesis and \textit{k}-anonymity based anonymizations override (almost) all identifying information in the image, we expect that these anonymizations cannot be reversed.
\textbf{E1.3:} For any other anonymization, we expect them to be partially reversible which means that reversal will result in higher accuracies than naive and parrot recognition but will not reach the clear data baseline.

The second aspect is about our purpose-built machine learning model that allows us to understand what makes reversibility possible.
\textbf{E2.1:} We expect that this model is able to at least partially reverse any anonymization that any other method can reverse. This would mean that the two processes responsible for reversal are actually reconstruction and inversion.
\textbf{E2.2:} We also expect our model to significantly outperform Pix2Pix for any global anonymization, namely Block Permutation and Pixel Relocation. Our model includes a linear layer that allows global inversion, i.e., not only anonymizations that perform the same transformation on all neighborhoods of pixels can be reversed.

The third aspect is the comparison of specialized de-anonymiza\-tions, general de-anonymizations, naive and parrot recognition.
\textbf{E3.1:} We expect that all general and specialized de-anonymizations result in higher accuracies than naive recognition and in many cases even parrot recognition. This is because a successful de-anonymization will result in individuals being more identifiable in the reversed images.
\textbf{E3.2:} We also expect that for any anonymization where a specialized de-anonymization exist, all general de-anonymizations can partially reverse the anonymization, but will generally have lower performance than the specialized de-anonymization. This is because specialized de-anonymizations can be specifically designed for the target de-anonymization while general approaches have to be anonymization-agnostic.

For the forth aspect, we investigate the generalizability of general de-anonymizations.
\textbf{E4.1:} When considering cases where training and test data were not anonymized using the exact same anonymization, we expect that identification accuracy decreases as parameters get less similar and do not expect de-anonymization to work at all if the anonymization method does not match.
\textbf{E4.2:} We expect that training the general de-anonymization on a different data set than the one used to test the anonymization will result in slightly lower identification accuracy, but will generally still work.

\subsection{Experiment Design}\label{sec:experiment_design}
To test \textbf{E1-3}, we perform re-identification experiments.
Initially, we generate a baseline by running our experiments without any anonymization or de-anonymization.
Then, for every anonymization which we introduced in the previous section, we test multiple configurations.
Like previous evaluation methodologies, we test naive and parrot recognition on the anonymized data without any de-anonymization.
We additionally test the reversibility evaluation methodology, both with any relevant specialized de-anonymizations as well as the general de-anonymizations Pix2Pix and our model.
The de-anonymizations tested for every anonymization can be found in Appendix~\ref{sec:combis}.
This sparse table once again highlights that general de-anonymization is needed to evaluate anonymizations because for many, no specialized approaches have been proposed.

For \textbf{E4}, we also conduct re-identification experiments.
We anonymize data using Gaussian Blur (kernel 29) and de-anonymize this data using models trained on images that were anonymized using Gaussian Blur (kernel 21, 25, 29, 33, 37), Gaussian Noise (sigma 200), DP Snow or Pixelation (size 16).
Additionally, we train both general de-anonymizations on one data set and then de-anonymize images of a different data set for all anonymizations.

\subsection{Data Sets}

We primarily use a subset of the commonly used CelebA data set \cite{liu_deep_2015} as the base set for our experiments.
We create our subset by sorting the identities in CelebA by their number of images and choose the top 5000 identities.
This is done because for the re-identification experiments, having more images per identity is preferential to allow for successful matching and to reduce the impact of outliers.
From those 5000 identities, we randomly choose 200 for our anonymization background set and 4800 for the evaluation data set of which 300 are for the test set and the remaining 4500 identities are used for de-anonymization training.
For our tests involving a different data set, we use a random subset of DigiFace-1M \cite{Bae_2023_WACV} with the same subset sizes as CelebA.  Both datasets contain a variation of face poses and accessories matching our social media scenario. More information about the used data sets can be found in Table~\ref{tab:data}.

Before our experiments, we run all images of our CelebA subset through a pre-processing pipeline.
Detecting a face bounding box is a first processing step of the majority of face-specific (de-) anonymizations and is usually performed by a state-of-the-art face detection algorithm that is not directly part of the actual (de-) anonymization.
Therefore, to improve performance and to remove any effects that degraded face detection on anonymized images may have on our results, we perform this face detection step once and then disable it whenever possible in subsequent methods.

\begin{table}[h!]
\centering
\caption{Properties of used data sets CelebA and DigiFace-1M}
\small
\begin{tabular}{p{3.0cm}|cc}
  & CelebA \cite{liu_deep_2015} & DigiFace-1M \cite{Bae_2023_WACV}\\
  \hline
  No. of identities & 10,177 & 110,000\\
  No. of images & 202,599 & 1,220,000\\
  No. of images in our subset & 136,485 & 140,000\\
  Avg. images per identity in our subset & 27.3 & 28\\
  Data origin & Celebrity images & Synthetic \\
  & from www & \\
\end{tabular}
\label{tab:data}
\end{table}

Our pre-processing is based on the pipeline of LightFace \cite{serengil_lightface_2020} and works as follows:
We use RetinaFace \cite{deng_retinaface_2019} to detect the bounding box of any faces in the images.
Note, that we do not use the bounding boxes provided by CelebA as we found them to be inaccurate at times and in order to resemble a standard face recognition pipeline more closely.
We choose the face with the largest bounding box that is fully in the image and crop the image to the smallest square that fully includes the face when rotated so that both eyes are on a horizontal line above the nose. Selecting the largest face in the image assumes a worst-case attacker, as the largest face in most cases contains the most identifying information and thus will be the most difficult to anonymize. Since anonymizations should protect people's identities even in worst-case scenarios, we consider this choice is reasonable for evaluating anonymizations.
These images are then resized to a resolution of 224x224 pixels which is the standard input size for LightFace.
We skip the pre-processing for DigiFace-1M as these synthetic images are already cropped to the face area.

\subsection{Evaluation Framework}

\begin{figure}[h!]
    \centering
    \includegraphics[width=0.4\textwidth]{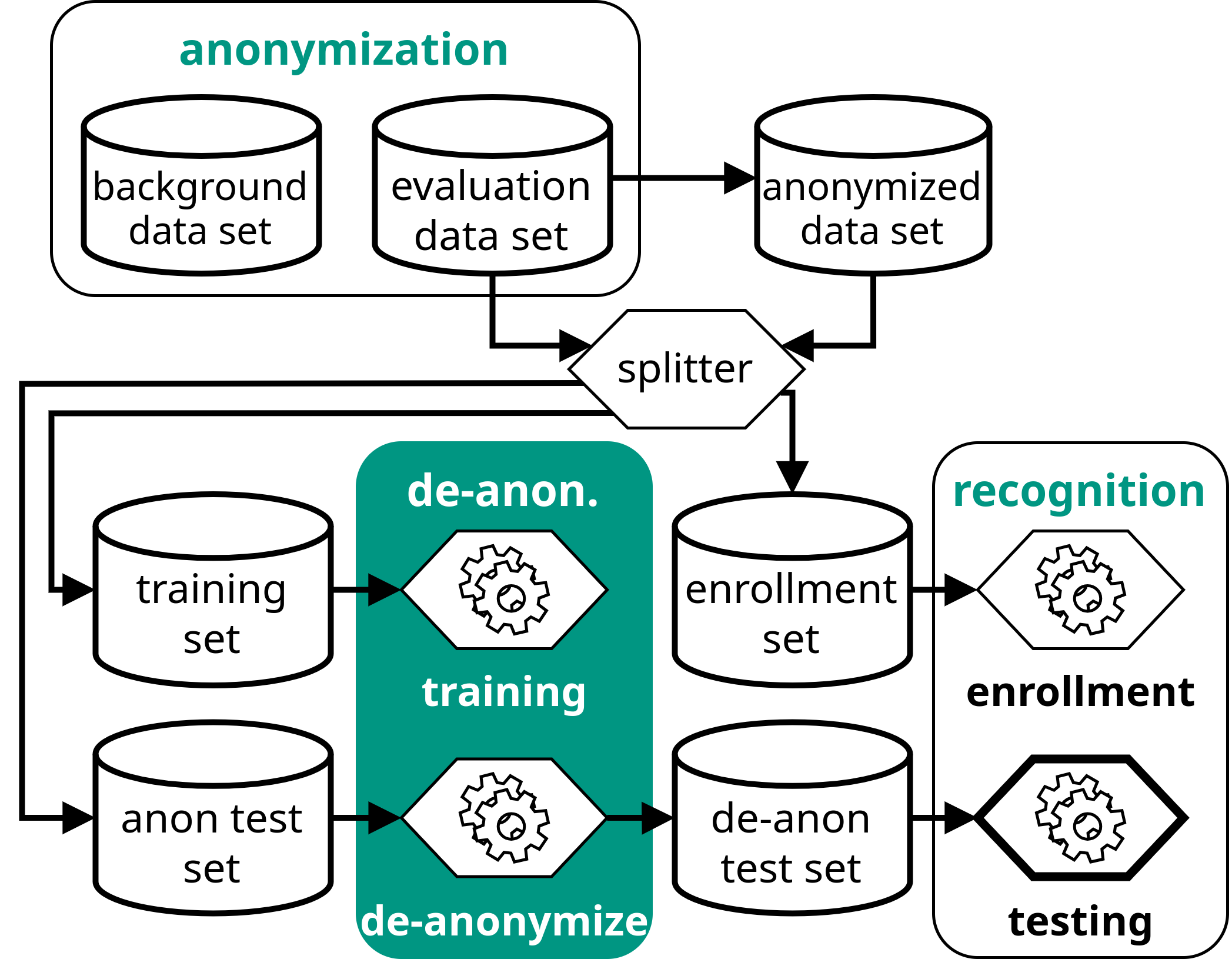}
    \caption{Data usage of our evaluation framework. The upper part depicts how the data sets are used for the anonymization, the bottom left show how the de-anonymization approach is trained and applied, and the right bottom shows how the recognition system is used.}
    \label{fig:systemdesign}
\end{figure}

In order to run our experiments, we implemented an evaluation framework (Figure \ref{fig:systemdesign} contains an overview of the data usage) that allows us to run the different experiments described in Section~\ref{sec:experiment_design}.

In a first step, the framework creates an anonymized data set.
For this our subset of CelebA or DigiFace-1M (already pre-processed) is split into evaluation data set and background data set.
Then the specified anonymization creates an anonymous copy of the evaluation data set, potentially using the background set.
Afterwards, evaluation and anonymized data set are disjointly split into training data set for the de-anonymization, the enrollment data set, and the test data set.
Enrollment and test share the same identities but contain different images while the training data contains all clear and anonymized images for all other identities.
The training set is used to train the de-anonymization before it is used to de-anonymize the test data set.
Finally, a face recognition method identifies the images in the de-anonymized test data set using the enrollment set and these results are used to calculate the metrics.
Because we have no clear indication which face recognition model may work the best on \mbox{(de-)} anonymized data, we test multiple.
We use pre-trained models of multiple state-of-the-art recognition models which are integrated into the LightFace framework: Facenet \cite{schroff_facenet_2015}, \mbox{VGGFace2} \cite{cao_vggface2_2018} and ArcFace \cite{deng_arcface_2019}.
Additionally, we use a combination of the face recognition model (fr-knn) \cite{geitgey_face_2021} which uses a pre-trained feature extractor based on \cite{king_dlib-ml_2009} and then classifies using k-nearest neighbours.
The framework was implemented in python (version 3.8) using the numpy (version 1.19.5) and scikit-learn (version 1.0.1) libraries.

\textit{Metrics.} Our goal is to measure the identifying information contained in de-anonymized images.
For this, our main metrics are the accuracies per identity of different face recognition models when performing re-identification experiments.
We then calculate the mean accuracy over all tested identities and a 95\%-confidence interval.

\textit{Parameters.}
Most anonymizations can be configured using parameters that determine their privacy-utility trade-off.
We choose these parameters based on common choices in related work or (if applicable) the method's author's recommendation.
We strive to evaluate the anonymizations on a realistic privacy-utility trade-off.
The specific parameters for each anonymization are included Appendix~\ref{sec:anonparams}.
Further parameters of de-anonymizations and the results of the hyperparameter search for our model can also be found in Appendix~\ref{sec:params}.

The code of our evaluation framework is available as part of an overall evaluation framework for biometric anonymization, which can be found at \textbf{\url{https://github.com/kit-ps/seba}}.

\subsection{Results}
\begin{figure*}[ht]\centering\setlength{\tabcolsep}{2pt}
\begin{tabular}{*{8}{>{\centering\arraybackslash}m{2.05cm}}}
  \includegraphics[width=2cm]{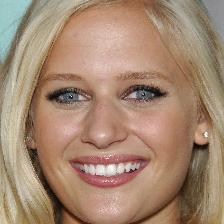} & \includegraphics[width=2cm]{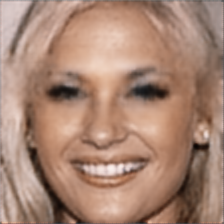} & \includegraphics[width=2cm]{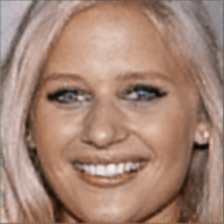} & \includegraphics[width=2cm]{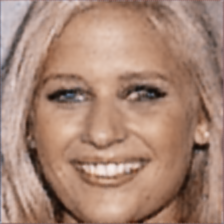} & \includegraphics[width=2cm]{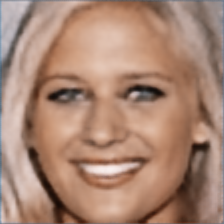} & \includegraphics[width=2cm]{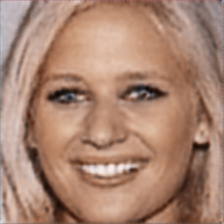} & \includegraphics[width=2cm]{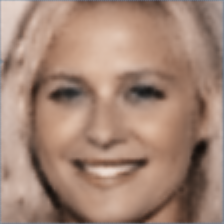} & \includegraphics[width=2cm]{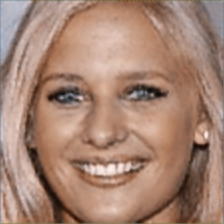}\\
  Clear & Eye Mask & Block Permut. & Pixel Reloc. & Gauss. Noise & Gauss. Blur & Pixelation & Fawkes\\
  \\
  \includegraphics[width=2cm]{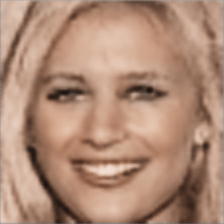} & \includegraphics[width=2cm]{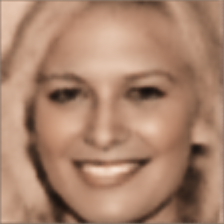} & \includegraphics[width=2cm]{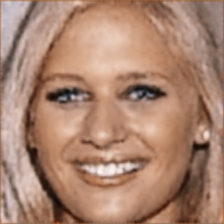}  & \includegraphics[width=2cm]{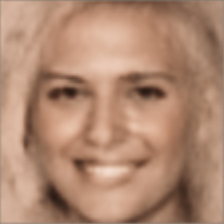} & \includegraphics[width=2cm]{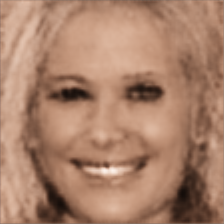} & \includegraphics[width=2cm]{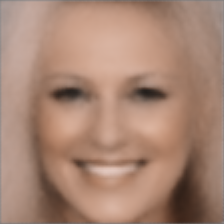} & \includegraphics[width=2cm]{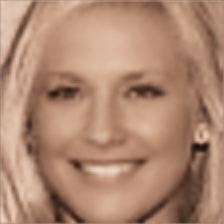} & \includegraphics[width=2cm]{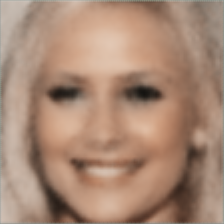}\\
  k-RTIO & DP Pix & DP Snow & DP Samp & \textit{k}-Same-Pixel & \textit{k}-Same-Eigen & DeepPrivacy & CIAGAN\\
\end{tabular}
\caption{De-anonymized images for different anonymization methods using our model}
\label{fig:deanonymization}
\end{figure*}

\begin{figure*}[h!]
    \includegraphics[width=\textwidth]{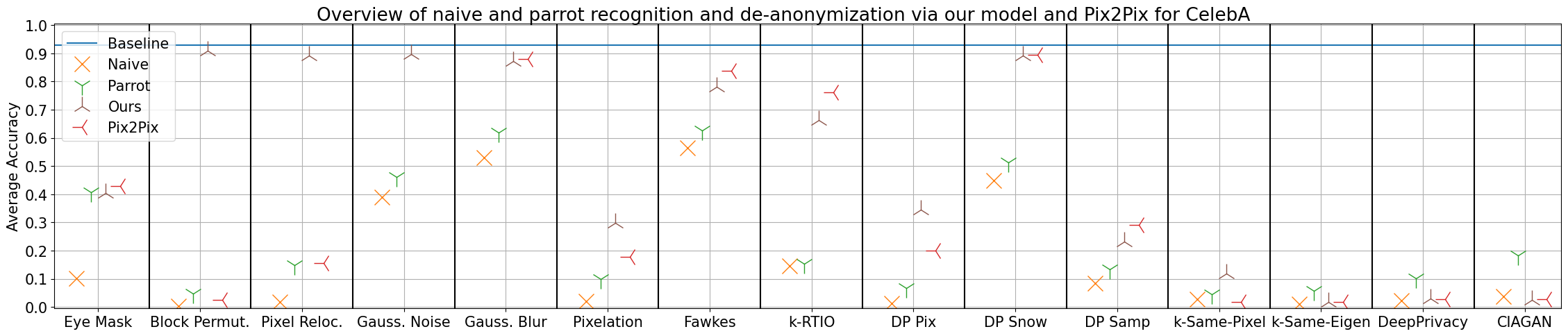}
    \caption{Average recognition accuracy for every anonymization method for baseline, naive, parrot, de-anonymized via our model, and de-anonymized via Pix2Pix; on 300 identities of CelebA.}
    \label{fig:res-overview}
\end{figure*}

\begin{figure}[h!]
    \includegraphics[width=\linewidth]{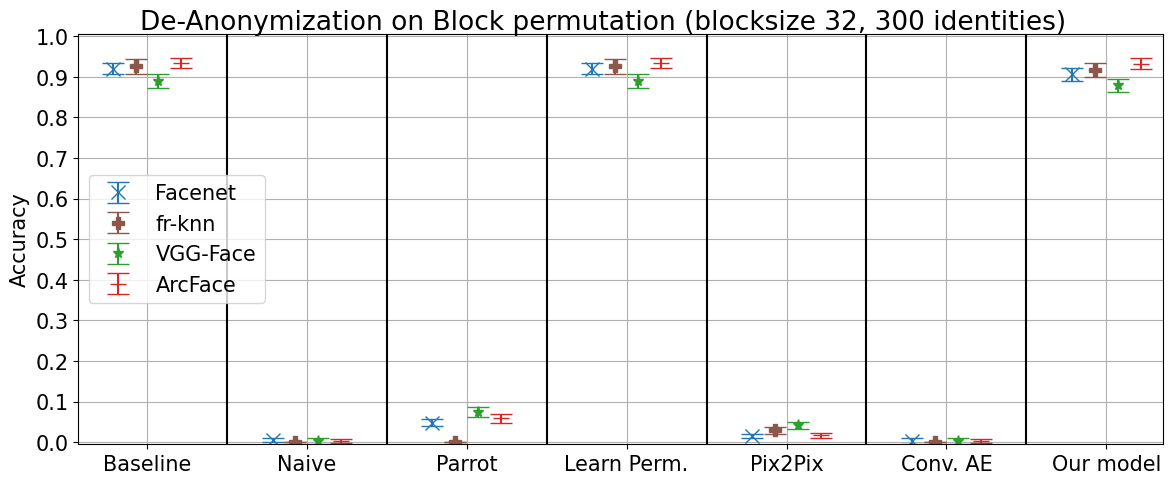}
    \caption{Recognition accuracy for Block Permutation (blocksize 32), given for baseline, naive, parrot, de-anon. via learn permutation, via Pix2Pix, via Conv. AE (our model without linear layer), and via our model; on 300 identities of CelebA.}
    \label{fig:res-blockpermutate-hl}
\end{figure}
\begin{figure}[h!]
    \includegraphics[width=\linewidth]{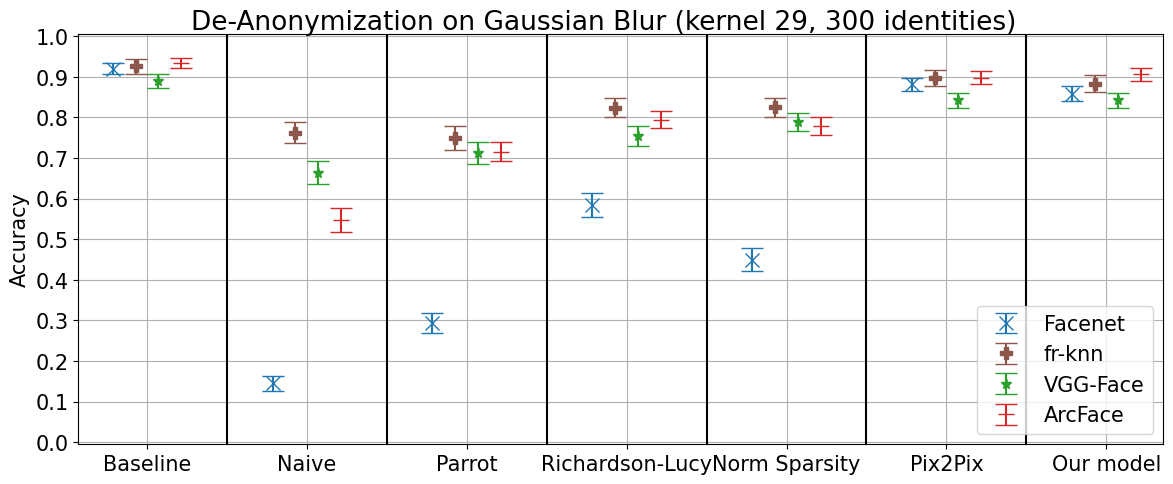}
    \caption{Recognition accuracy for Gaussian Blur (kernel 29), given for baseline, naive, parrot, de-anon. via bicubic interpolation, via Richardson-Lucy, via Norm-Sparsity, via Pix2Pix, and via our model; on 300 identities of CelebA.}
    \label{fig:res-blur-hl}
\end{figure}
\begin{figure}[h!]
    \includegraphics[width=\linewidth]{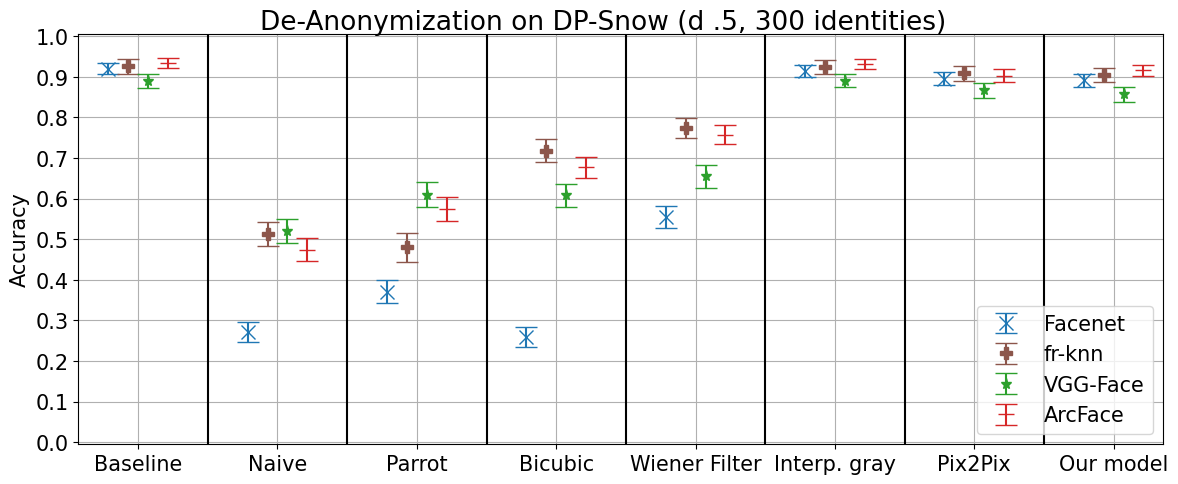}
    \caption{Recognition accuracy for DP Snow, given for baseline, naive, parrot, de-anon. via bicubic interpolation, via Wiener Filter, via interpolate gray, via Pix2Pix, and via our model; on 300 identities of CelebA.}
    \label{fig:res-dpsnow-hl}
\end{figure}
\begin{figure}[h!]
    \includegraphics[width=\linewidth]{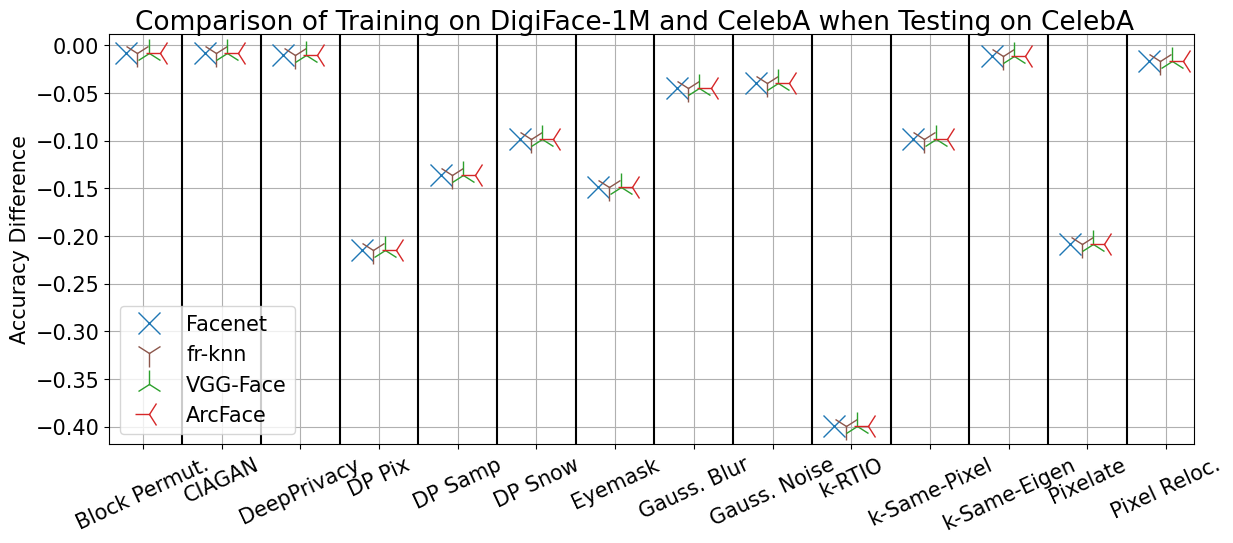}
    \caption{Difference in recognition accuracy when our model is trained on DigiFace-1M instead of CelebA for all anonymizations; tested on 300 identities of CelebA.}
    \label{fig:res-cmp-digiface}
\end{figure}
\begin{figure}
    \includegraphics[width=\linewidth]{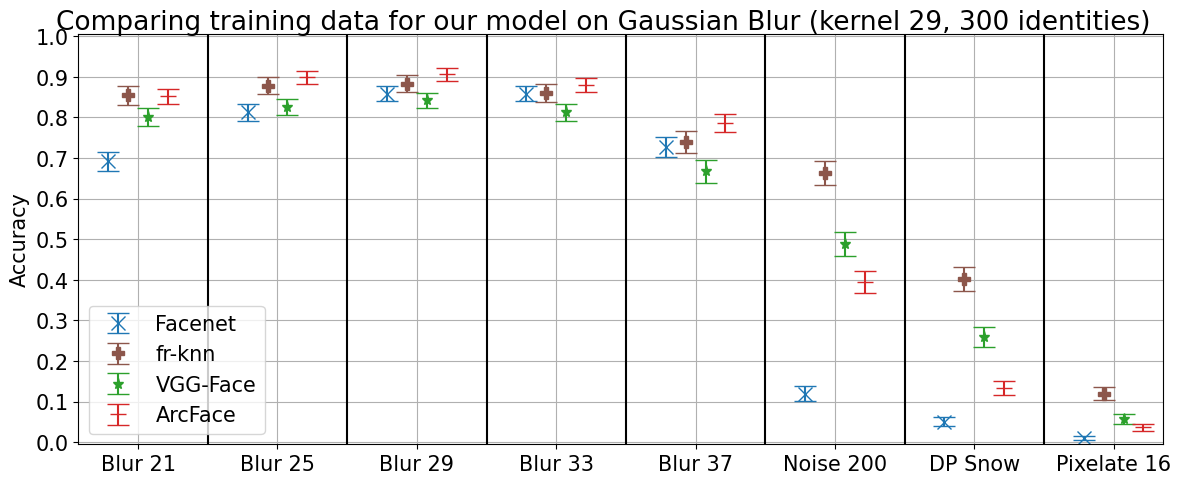}
    \caption{Recognition accuracy for Gaussian Blur (kernel 29), given for de-anon. via our model using Gaussian Blur (Kernel 21, 25, 29, 33, 37), Gaussian Noise (sigma 200), DP Snow, or Pixelation (size 16) as training data for our de-anon. model; on 300 identities of CelebA.}
    \label{fig:res-cmp-models}
\end{figure}

In the following, we first present our general findings as illustrated by the selection of figures in this section. Afterwards, we consider each of our expectations and to what extent we find evidence for them in our results.

\subsubsection{General Findings}
Example images that were de-anonymized using our model can be found in Figure~\ref{fig:deanonymization}.
An overview over the results for naive and parrot recognition, Pix2Pix, and our model can be found in Figure~\ref{fig:res-overview} for CelebA and Figure~\ref{fig:res-overview-digiface} for DigiFace-1M.
This plot includes the results for the different experiments for all anonymizations.
Generally, a high value indicates that the experiment showed that the images still contained enough personal information to identify an individual.
Therefore a successful anonymization would result in low values for all experiments.
For naive and parrot recognition as well as Pix2Pix and our model, the average of all recognition models is shown.
We find that for many anonymizations accuracies exceeding 50\% can be measured and that for most anonymizations, Pix2Pix and our model results in significantly higher values than the other experiments.

This section also includes plots for the anonymizations Block Permutation (Figure~\ref{fig:res-blockpermutate-hl}), Gaussian Blur (Figure~\ref{fig:res-blur-hl}) and DP Snow (Figure~\ref{fig:res-dpsnow-hl}).
Plots with all de-anonymizations as well as plots for all other anonymizations can be found in Appendix~\ref{sec:fresults}.
In these plots, different de-anonymizations are compared against the baseline of clear data as well as naive and parrot recognition on the anonymized data.
High differences between naive or parrot recognition and a de-anonymization indicate that this method was able to reverse the anonymization and to re-create an image on which face recognition methods were able to identify an individual.
For Block Permutation, we find that the specialized learn permutation de-anonymization is able reach clear level performance with our model not much lower. Pix2Pix however is not able to reverse Block Permutation, similarly to a version of our model without linear layer. This highlights inversion as the main underlying principle enabling reversibility for Block Permutation.
While the specialized approaches for Gaussian Blur are able to increase identification accuracies over naive and parrot level, they are below our model and Pix2Pix which also do not reach clear level.

In Figure~\ref{fig:res-cmp-digiface}, the effects of training our model on DigiFace-1M and testing on CelebA can be seen. For each anonymization the accuracies of our model when trained on CelebA are compared to when training on DigiFace-1M. We find that for most anonymizations our model generalizes well.
In Figure~\ref{fig:res-cmp-models} for our model and Figure~\ref{fig:res-cmp-models-p2p} for Pix2Pix, the effects of the training data not exactly matching the testing data can be seen.
In all cases, the test data was anonymized using Gaussian Blur (kernel 29), the general de-anonymization was however trained with data that was anonymized using other anonymizations.
We find that the best result is achieved when training and testing data match and the performance decreases as the two data sets get less similar.

\subsubsection{Evaluating our Expectations}
Our results in Figure~\ref{fig:res-overview} indicate that \textbf{E1.1-3} are true.
We find that face recognition on images de-anonymized using our model performs at almost baseline performance for permutations (\textbf{E1.1}).
For none of the anonymizations in the categories synthesis and \textit{k}-Anonymity does any identification accuracy of our model exceed 15\% (\textbf{E1.2}), see Figure \ref{fig:res-overview}, see \textit{k}-Same-Pixel (Figure \ref{fig:res-ksamepixel}), \textit{k}-Same-Eigen, DeepPrivacy (Figure \ref{fig:res-deepprivacy}) and CIAGAN (Figure \ref{fig:res-ciagan}).
For all other anonymizations, we find as predicted in \textbf{E1.3} that they can be partially reversed and de-anonymizations thereby increase re-identification and similarity metrics.
We do find that the level of de-anonymization varies significantly between anonymizations from close to perfect for DP Snow (Figure \ref{fig:res-dpsnow-hl}) to very little improvement in DP Samp (Figure \ref{fig:res-dpsamp}).

We also find evidence for \textbf{E2.1-2}.
Especially, we find all anonymi\-zations with specialized de-anonymizations according to Table~\ref{tab:experiments}, are categorized as either partially or highly reversible in Figure~\ref{fig:res-reversibility}. This means that no specialized de-anonymization used a different reversal processes than the two on which our model is built, otherwise it would not have been able to reverse these anonymizations (\textbf{E2.1}).
For the permutations (see Figure~\ref{fig:res-blockpermutate-hl} and Figure~\ref{fig:res-pxlreloc}), we see large differences between Pix2Pix and our model as expected (\textbf{E2.2}), confirming that the linear layer that is exclusive to our model is necessary to handle global inversions.
We also find evidence for the necessity of the linear layer in our model by comparing its result to a version without the linear layer.
The identification accuracy is increased by adding the linear layer particularly for the permutation-based anonymizations (see Figure~\ref{fig:res-blockpermutate-hl} \&~\ref{fig:res-pxlreloc}), but also for DP Pix, Eye Masking, k-RTIO and Pixelation (see Figures~\ref{fig:res-dppix},~\ref{fig:res-eyemask},~\ref{fig:res-krtio} \&~\ref{fig:res-pixelate16}).

When considering \textbf{E3.1}, we find that as expected all de-anonymi\-zations achieve higher accuracies than naive recognition, see Figure~\ref{fig:res-overview}.
However, there are some exceptions in which parrot recognition performs better than our model and even naive recognition is within a small margin, namely Eye Masking, DeepPrivacy, CIAGAN and \textit{k}-Same-Eigen.
All these anonymizations share that they remove large parts of the face  and the only option for a de-anonymization therefore is to reconstruct these areas using the general structure of faces.
We see this as an indication that no information is better for the face recognition than incorrectly reconstructed information.
We also find \textbf{E3.2} to be often correct, however also find cases in which the general de-anonymizations match or even outperform specialized approaches.
We attribute this to the general approaches being trained on only face images while specialized approaches often may also be used on non-face images.

Our results in Figures~\ref{fig:res-cmp-models}~\&~\ref{fig:res-cmp-models-p2p} and Figure~\ref{fig:res-cmp-digiface} also indicate the correctness of \textbf{E4.1} and \textbf{E4.2}, respectively.

\subsection{Why are Some Anonymizations Reversible?}

Based on our findings, we present here what we consider the main reasons why anonymizations are vulnerable to reversal.

\subsubsection{The identifying information has only been obfuscated and not removed}
Some approaches only generalize the information contained in the image by averaging it, for example Pixelation, Gaussian blur, or DP Samp. We find these approaches to be partially reversible because the identifying information is only obfuscated and not removed, making them susceptible to reconstruction of the original information. Furthermore, only shuffling the identifying information in the image, as Block Permutation and Pixel Relocation do, is susceptible to inversion and therefore allows the anonymization to be completely reversed. In comparison, the anonymizations that remove or override the identifying information in the face are the least reversible, DeepPrivacy and CIAGAN both generate new faces and thus effectively remove the identifying information from the face in the image. A similar effect can be seen with the \textit{k}-anonymity based approaches, which also overwrite the identifying information with the data of other faces.

\subsubsection{Not all identifying information has been anonymized}
Anonymizations that consider only parts of the face are reversible, as Eye Mask shows. So focusing only on parts of the face is not enough to remove its identification potential. The same problem can be observed with the anonymizations that apply random noise to the images (DP Snow and Gaussian Noise). Since these techniques do not change every pixel, the overall characteristics of the face remain intact and the original image can be reconstructed using the unchanged pixels and the learned general knowledge about face anatomy (see Figure~\ref{fig:deanonymization} the reconstructed eye section for Eye Mask).
This means that all parts of a face contain identifying information. This is reinforced by the knowledge that earlier not-machine learning face recognition approaches used proportions of the entire face to identify individuals \cite{jafri_survey_2009}.

\subsubsection{Reliance on potentially unsuitable formal guarantees}
We tested multiple anonymizations that are based on methods proven to fulfill the notion of DP.
However, our general de-anonymization attacker was still able to partially reverse these anonymizations
While this could be due to our adaptation for RGB images, it could possibly also be a result of the proof's inability to capture the real-world problem.
For example the proof for DP Snow defines two neighboring images as differing in one pixel and not as showing two different identities. %
While our attacker cannot recover the exact color of removed pixels, the identity remains recoverable. Further, DP assumes that the data is uncorrelated, which for pixels in an image is not the case.
This highlights that blindly relying on formal guarantees might be dangerous without an additional empirical evaluation in a more realistic scenario.

\section{Utility}\label{sec:userstudy-utility}
After our exhaustive investigation into reversibility, we now want to consider utility in order to make conclusions about which anonymizations to use in practice.
After all, a complete evaluation of any anonymization requires both privacy and utility to be evaluated.
We considered both a human-centric evaluation via a user study and a computational evaluation via similarity metrics.
In this section we will focus on the the design and results of our user study.
The utility evaluation using computational metrics can be found in Appendix~\ref{sec:comp-utility}.

The utility goals that anonymizations try to satisfy are diverse and difficult to compare.
We therefore focus on utilities that fit our data publishing scenario well.
This means that when users upload an image to a social media site in which either they themselves or bystanders are anonymized, they want this picture to still be visually appealing and/or appear natural.
To evaluate to which extent the considered anonymizations fulfill this goal, we design a user study adapted from previous work by Hasan et al., Cyr et al. and Li et al. \cite{hasan_can_2019, cyr_exploring_2009, li_effectiveness_2017}.

\begin{figure*}
    \includegraphics[width=\textwidth]{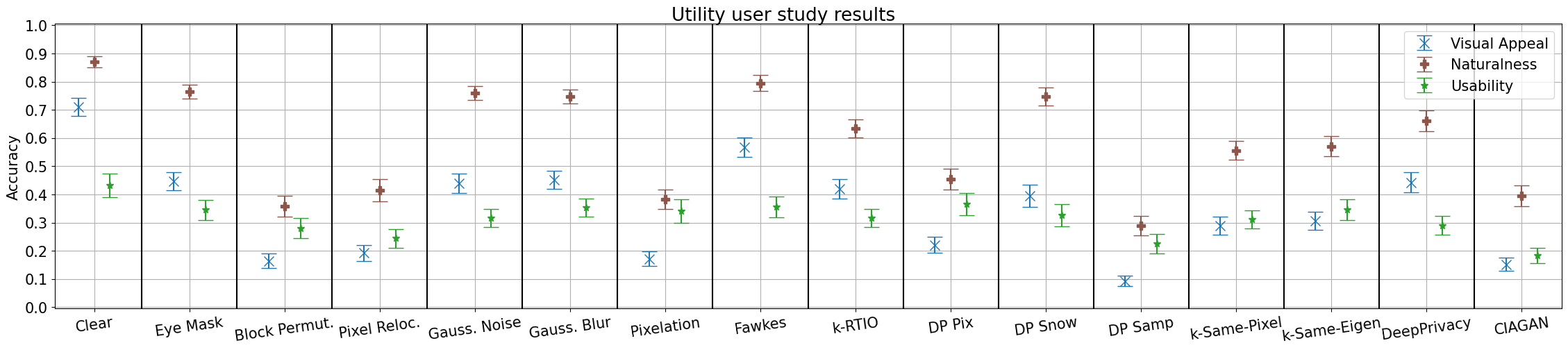}
    \caption{User study agreement scores for our three statements (visual appeal, naturalness and usability) for clear and all anonymizations.}
    \label{fig:res-userstudy-utility}
\end{figure*}

We randomly select 10 images (five male, five female) from the CelebA dataset which we anonymize with each of our 15 anonymizations.
Participants are shown each picture exactly once with a random anonymization (including none) though any anonymization may only appear once per participant.
Participants are asked via a seven point Likert scale from strongly disagree (0.0) to strongly agree (1.0) if they agree with three statements: (1) The picture is visually appealing. (2) The picture shows a natural human being. (3) I would use this anonymization when using social media.
We recruit 505 participants (limited to individuals who regularly use social media), which results in an average of 28.1 (standard deviation 5.2) votes per image.
We excluded 55 participants because of failed attention checks, which results in 450 participants (205 female, 244 male, 1 preferred not to say) with an average age of 33.2 (standard deviation 11.1) years.
In Figure~\ref{fig:res-userstudy-utility} we show the mean over all votes for each anonymization and a 95\% confidence interval.

When comparing the three statements, we find them to generally correlate.
However, naturalness scores usually higher than visual appeal and usability does not exceed 0.45 as a maximum or 0.15 as a minimum.
This indicates that the overall willingness of participants to use anonymizations (even when they are imperceptible in the case of clear) is not very high, even though over 60\% of participants agreed with the statement "I am worried about automatic face recognition on social media."
Visual appeal could be limited by the resolution of the images and the crop to the face region which might be negatively perceived for a social media scenario.
This could potentially also be improved by showing the participants anonymized images of themselves.
For these reasons, we focus on naturalness for our further analysis.
When comparing anonymizations, we find as expected clear imags to achieve the highest scores.
Fawkes also achieves high scores which is not surprsing considering it is designed to be imperceptible.
Commonly used methods such as Eye Masking, Noise and Blurring score above average which could be due to them being familiar to users.
\begin{figure}
    \includegraphics[width=\linewidth]{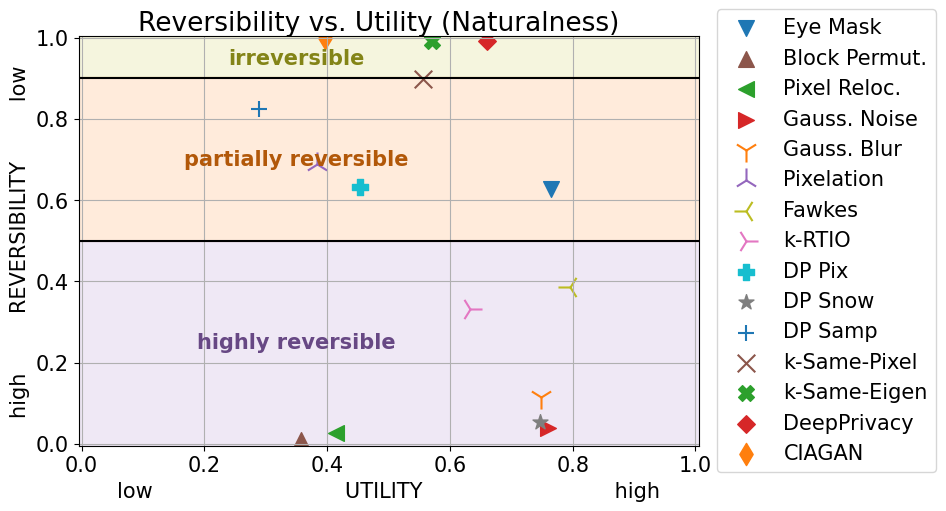}
    \caption{Reversibility over Utility (here: naturalness) for all anonymizations on CelebA and a categorization of anonymizations in irreversible, partially reversible, and highly reversible.}
    \label{fig:res-reversibility}
\end{figure}

As a takeaway, we  create a reversibility metric and plot it over the naturalness utility values in Figure~\ref{fig:res-reversibility}.
\textbf{Reversibility} is here defined as the average accuracy improvement of de-anonymization over naive recognition compared to clear level.
This allows us to categorize anonymizations into irreversible, partially reversible, and highly reversible. \\
We find that while Fawkes clearly provides the most utility, it is also highly reversible.
Of the anonymizations with above average (0.57) naturalness, DeepPrivacy is the only one that achieves low reversibility with the \textit{k}-Anonymity-based methods closely behind.
While one might generally except a trade-off between reversibility and utility, this is indeed not what we find.
Instead, some of the anonymizations that are partially (or even highly) reversible, also do not provide high levels of utility while anonymizations that are irreversible in some cases are also able to provide decent utility.
This indicates that the privacy-utility trade-off that is generally expected for easy-to-parameterize anonymizations (such as noise injection) does not seem to apply to the comparison between different anonymization methods. %
It is also important to note that privacy and reversibility are not the same, although they are closely related.
Nevertheless, this plot allows for a direct comparison of the reversibility of anonymizations (given specific configuration parameters).

\section{Ethical Considerations}\label{sec:ethics}
\makeatletter\if@ACM@anonymous\makeatother
The user study data collection was approved by our university's institutional review board and was conducted in accordance with the Declaration of Helsinki.
\else
The user study data collection was approved by the ethics commission of the Karlsruhe Institute of Technology (research project "Evaluierung von Gesichtsanonymisierungen") and was conducted in accordance with the Declaration of Helsinki.
\fi
All data was collected as an anonymous online survey in Feburary 2024 using an online recruitment platform\footnote{https://www.prolific.com/}. Participation took a median of four minutes and participants were paid an average of \pounds12.68 per hour.

Responsible disclosure: As all of the tested evaluations in this paper have been selected from the scientific literature there are no vendors or specific anonymization services, that we are aware of, that can be contacted to disclose our findings to.

\section{Limitations}\label{sec:limitations}
While we consider the simplicity of our machine learning model's design to be a feature which allows a better understanding of its results, we acknowledge that a more advanced model might result in even better de-anonymization results.
Improvements like this may include further tuning of hyperparameters, longer training with more data or improvements to the data pre-processing.
Further, we consider that SSIM as a loss function does not ideally capture our goal which is the reconstruction of the identity in the image, not the pixel values.
Therefore, an identity loss function rather than an image similarity loss function like SSIM could be more suitable. %
Overall, we think that these limitations are negligible and do not diminish our conclusions.

\section{Conclusion and Future Work}\label{sec:conclusion}

There are significant privacy risks associated with the collection of biometric data which facilitates the requirement for anonymizations.
Face anonymizations are commonly evaluated using a weak attacker model without considering reversibility.
At the same time, strong attackers for specific anonymizations have been shown to be successful and general de-anonymizations have been shown to be feasible.
An in-depth understanding of face anonymization reversibility was however still missing.
In this work, we investigate this phenomenon exhaustively by considering different aspects and conducting a large number of experiments.

We find that a majority of anonymization methods is at least partially reversible and therefore protects the privacy of individuals less than previously thought, at least under our parameter choices.
Our general de-anonymization is able to successfully reverse anonymized images in 11 out of 15 cases.
In comparison to the common methodology, and often even specialized approaches, the general de-anonymizations result in significantly increased identification accuracy.
This highlights the need for strong attacker models when evaluating anonymizations.
When considering what makes reversal possible, we find that the underlying processes are reconstruction and inversion.
We find that while trained general de-anonymization also work on other data sets, when the anonymization method does not match between training and test data, results suffer significantly.
Finally, considering the utility of anonymizations, we find that in general, there does not seem to be a reversibility-utility trade-off between different anonymizations, but rather anonymizations can be both irreversible and provide decent utility.

We also analyze what causes anonymizations to be reversible.
Based on this, takeaways can be derived for future anonymization designers.
Irreversible anonymizations should remove and replace identifying information in the data, as obfuscations can be reconstructed or inverted.
Also, all identifying information must be anonymized because any remaining information might be used to reconstruct the data.
Finally, while formal guarantees might allow for a better quantification, it should be considered good practice to add an empirical evaluation.

In conclusion, in this work, we have conducted an exhaustive investigation of face anonymization reversibility in order to understand how and when reversal is possible.
This understanding will help construct anonymizations that are actually irreversible and thereby better protect the privacy of individuals in the future.

\begin{acks}
This work was funded by the Topic Engineering Secure Systems of the Helmholtz Association (HGF) and supported by KASTEL Security Research Labs, Karlsruhe.
Funded by the German Research Foundation (DFG, Deutsche Forschungsgemeinschaft) as part of Germany’s Excellence Strategy – EXC 2050/1 – Project ID 390696704 – Cluster of Excellence “Centre for Tactile Internet with Human-in-the-Loop” (CeTI) of Technische Universität Dresden.
\end{acks}

\bibliographystyle{ACM-Reference-Format}
\bibliography{main}

\appendix

\section{Anonymization Parameters}
\label{sec:anonparams}
Table \ref{tab:eval-params} shows the used parameters for all anonymization methods.
Methods that do not have any configurable parameters are excluded.

\begin{table}[h!]
\centering
\caption{Parameters of anonymization methods}
\small
\begin{tabular}{l|l}
  Method & Parameters\\
  \hline
  Block Permutation & block size 32\\
  Pixel Relocation & steps 50\\
  Gaussian Noise & $\sigma$ 200\\
  Gaussian Blur & kernel size 29\\
  Pixelation & size 16\\
  Fawkes & mode high\\
  DP Pix & $\epsilon$ 5, $b$ 12, $m$ 16\\
  DP Snow & $\delta$ 0.5\\
  DP Samp & $\epsilon$ 25, $k$ 24, $m$ 12\\
  \textit{k}-Same-Pixel & $k$ 10 \\
  \textit{k}-Same-Eigen & $k$ 10\\
\end{tabular}
\label{tab:eval-params}
\end{table}

\section{De-anonymization Parameters}
\label{sec:params}
For all specialized de-anonymizations as well as Pix2Pix, we use the standard hyperparameters recommended by their authors.
For our model, we conduct a hyperparameter search by running a variety of configurations testing de-anonymization performance on a data set anonymized using Gaussian Blur (kernel 29).
For each considered hyperparameter, we test multiple options and choose the one that results in the best performance in this experiment.
We choose LeakyReLU as our activation function, SSIM as our loss function, an initial learning rate of 0.0001 and a batch size of 64.
We were able to further improve results by adding a reduce-on-plateau learning rate adaption to our training that multiplies the current learning rate by 0.75 if the validation loss does not improve for five epochs.
We train for a maximum of 200 epochs but stop early when we do not measure an improvement in validation loss for 20 epochs.

\section{(De-) Anonymization Combinations}
\label{sec:combis}
Table~\ref{tab:experiments} shows the tested de-anonymizations for every anonymization in this paper.
\begin{table}[h!]
\centering
\caption{Combinations of anonymization and de-anonymi\-zation methods evaluated as part of this paper}
\small\setlength{\tabcolsep}{2.5pt}
\begin{tabular}{l|*{12}{>{\raggedright\arraybackslash}m{0.365cm}}}
  Method & \rotf{Linear/Bicubic} & \rotf{Wiener Filter} & \rotf{Richardson-Lucy} & \rotf{Wavelet Denoising} & \rotf{DIC SR} & \rotf{Norm Sparsity} & \rotf{Stripformer} & \rotf{MPRNet} & \rotf{Neighbor Interpol.} & \rotf{Learn Permutation} & \rotf{Pix2Pix} & \rotf{\textbf{our model}}\\
  \hline
  Eye Mask & & & & & & & & & & & \cmark & \cmark\\
  Block Permut. & & & & & & & & & & \cmark & \cmark & \cmark\\
  Pixel Reloc. & & & & & & & & & & \cmark & \cmark & \cmark\\
  Gauss. Noise & \cmark & \cmark & \cmark & \cmark & & & & \cmark$^{\mathrm{a}}$ & & & \cmark & \cmark\\
  Gauss. Blur & \cmark & \cmark & \cmark & & & \cmark & \cmark & \cmark$^{\mathrm{b}}$ & & & \cmark & \cmark\\
  Pixelation & \cmark & \cmark & & & \cmark & & & & & & \cmark & \cmark\\
  Fawkes & \cmark & \cmark & \cmark & \cmark & & & & \cmark$^{\mathrm{a}}$ & & & \cmark & \cmark\\
  DP Pix & \cmark & \cmark & \cmark & & & & & \cmark$^{\mathrm{a}}$ & & & \cmark & \cmark\\
  DP Snow & \cmark & \cmark & \cmark & \cmark & & & & \cmark$^{\mathrm{a}}$ & \cmark & & \cmark & \cmark\\
  DP Samp & & & & & & & & & & & \cmark & \cmark\\
  \textit{k}-Same-Pixel & & & & & & & & & & & \cmark & \cmark\\
  \textit{k}-Same-Eigen & & & & & & & & & & & \cmark & \cmark\\
  DeepPrivacy & & & & & & & & & & &\cmark & \cmark\\
  CIAGAN & & & & & & & & & & & \cmark & \cmark\\
  k-RTIO & & & & & & & & & & & \cmark & \cmark\\
\end{tabular}\\
\begin{footnotesize}\raggedright
  $^{\mathrm{a}}$ Denoising  $^{\mathrm{b}}$ Deblurring\\
\end{footnotesize}
\label{tab:experiments}
\end{table}

\section{Human Evaluation of Reversibility}\label{sec:userstudy}
We conduct a user-study to test whether machine learning face recognition on de-anonymized images can identify individuals better than human observers on anonymized images.
Both McPherson et al. \cite{mcpherson_defeating_2016} and Hao et al. \cite{hao2020robustness} claim that their results indicate that humans are no longer the gold standard for evaluating the effectiveness of anonymizations.
However they don't provide any evidence for this claim.
As we want to investigate reversibility and thereby its impacts on machine learning face recognition as well as human observers, we therefore conduct this experiment.

The experiments in the previous section use a large number of images in the enrollment set (ca. 6000) which means that this experiment design is not feasible for a user study.
We therefore opt to conduct face verification experiments where participants decide whether two face images, one of which is anonymized, show the same person.
The rationale is that if a participant is not able to recognize that two images belong to the same person, the anonymization was successful at protecting this person's identity.

\begin{figure}[h!]
    \includegraphics[width=\linewidth]{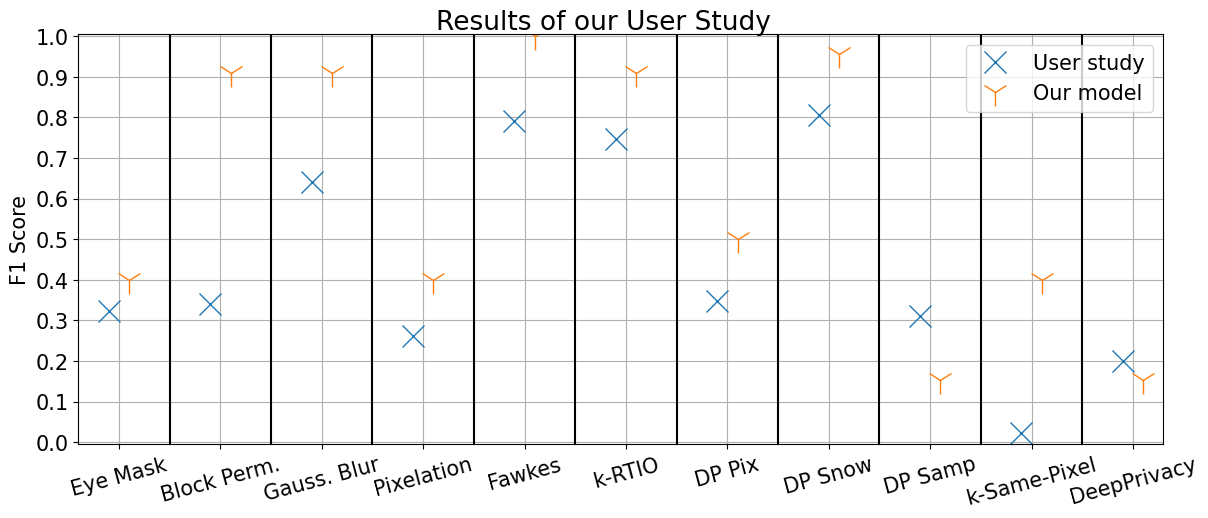}
    \caption{F1 Scores of the user study and face verification on de-anonymized images via our model for eleven anonymizations.}
    \label{fig:res-survey}
\end{figure}

To reduce the scope of the study, we do not test anonymizations that are very similar to others while making sure to include an anonymization from every category.
For each anonymization, we randomly choose 8 pairs of images of which 4 pairs show women and 4 pairs show men (as determined by the CelebA attributes) and of which 4 pairs show the same person in both images and 4 do not.
For each pair of images, we ask participants (n=98; minimum number of votes per image pair=34) whether both images show the same person and they can answer 'yes', 'no' or 'not sure'.
We also use three machine learning models (Facenet, VGG-Face2 and ArcFace) to verify the identity on the same chosen images, using the de-anonymized image via our model instead of the anonymized image.
For each anonymization, we calculate the F1-Score over all responses/models on all matching image pairs.

The results of our user study are shown in Figure~\ref{fig:res-survey}.
We find the expectation that machine learning outperforms human observers to be generally correct.
However, for the vast majority of anonymizations, the scores are very close to each other, within 0.15.
The slightly lower scores of user study could be attributed to a tendency to choose the 'not sure' option which is not available to the machine learning models.
Block Permutation is the only anonymization where face recognition performs significantly better which could be a result of it changing the overall structure of the image.
The only cases of human observers performing better are DP Samp and DeepPrivacy, which, however, have low scores in both cases.

Our results show that the combination of reversing anonymization and then performing face recognition generally outperforms human observers.

We acknowledge that our user study could be improved by collecting more data.
Both more participants and more samples per anonymization could be used to further increase confidence in our results.
In the current form, the random choice of images per anonymization leads to high variances when the identity decision is already difficult on clear images.

The user study data collection was approved by our university's institutional review board and was conducted in accordance with the Declaration of Helsinki.
All data was collected as an anonymous online survey in October 2022. The recruitment was done by advertising on social media, via email, and through direct recruitment of colleagues and friends. We did not collect socioeconomic data or pay compensation.

\section{Computational Utility Evaluation}\label{sec:comp-utility}

\begin{figure*}
    \includegraphics[width=\textwidth]{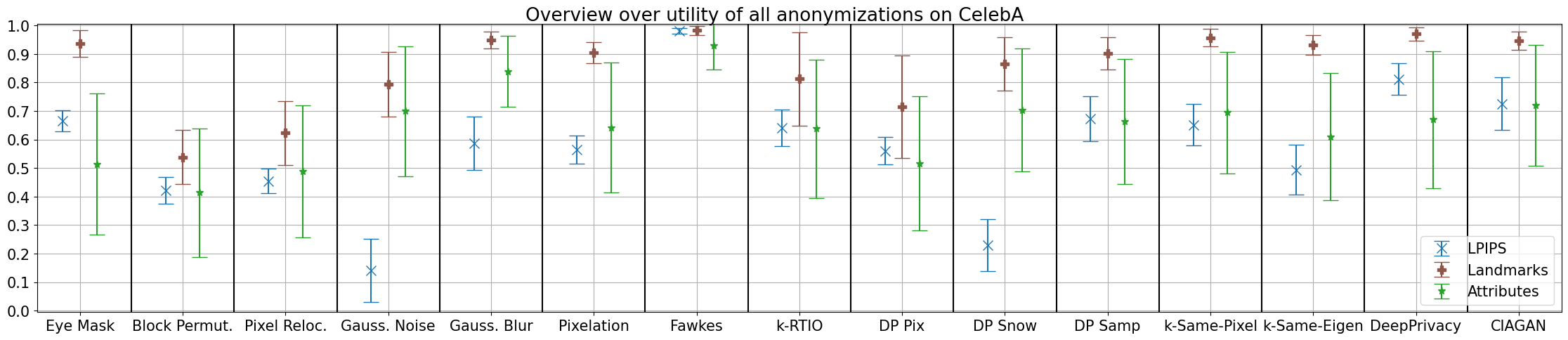}
    \caption{Mean utility and 95\% confidence interval via perceptual similarity (LPIPS), attribute similarity, and landmark similarity for all anonymizations; on 300 identities of CelebA.}
    \label{fig:res-comp-utility}
\end{figure*}

Besides our human-centric evaluation, here we perform a utility evaluation using computational similarity metrics.
We consider three main goals. The first is for the anonymized image to be similar to the original, the second is to preserve the face attributes, and the third is to preserve the landmark locations of the face.

We measure the similarity of the images by using the Learned Perceptual Image Patch Similarity (LPIPS)~\cite{Zhang_2018_CVPR}.
For attribute similarity, we calculate the mean absolute error between attributes (sex, race, emotion and age) recognized by DeepFace~\cite{serengil_lightface_2020} (using rooted mean squared error on sub-attributes where necessary).
For landmark similarity we used the mean Euclidean distance of the six keypoints detected by Google AI's mediapipe's face detector~\cite{mediapipe}.
We normalize all three metrics between 0 and 1 using the absolute minimum and maximum recorded for any image for any anonymization. Higher values always indicate better utility.

The results of these three metrics for all our anonymizations can be found in Figure~\ref{fig:res-comp-utility}.
We find that all scores are fairly high and surprisingly find the three metrics to be similar for most anonymizations.
The notable exception being noise-based anonymizations achieving significantly lower perceptual similarity.\\

\section{Further Results}
\label{sec:fresults}
\begin{figure}[h!]
    \includegraphics[width=\linewidth]{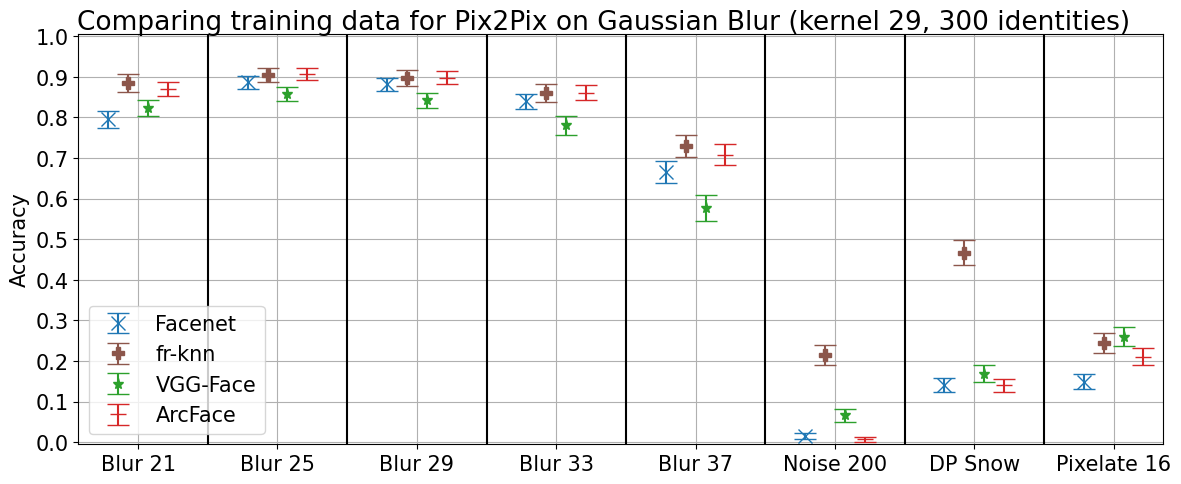}
    \caption{Recognition accuracy for Gaussian Blur (kernel 29), given for de-anon. via Pix2Pix trained on Gaussian Blur (Kernel 21, 25, 29, 33, 37), Gaussian Noise (sigma 200), DP Snow, or Pixelation (16)}
    \label{fig:res-cmp-models-p2p}
\end{figure}

\begin{figure}[h!]
    \includegraphics[width=\linewidth]{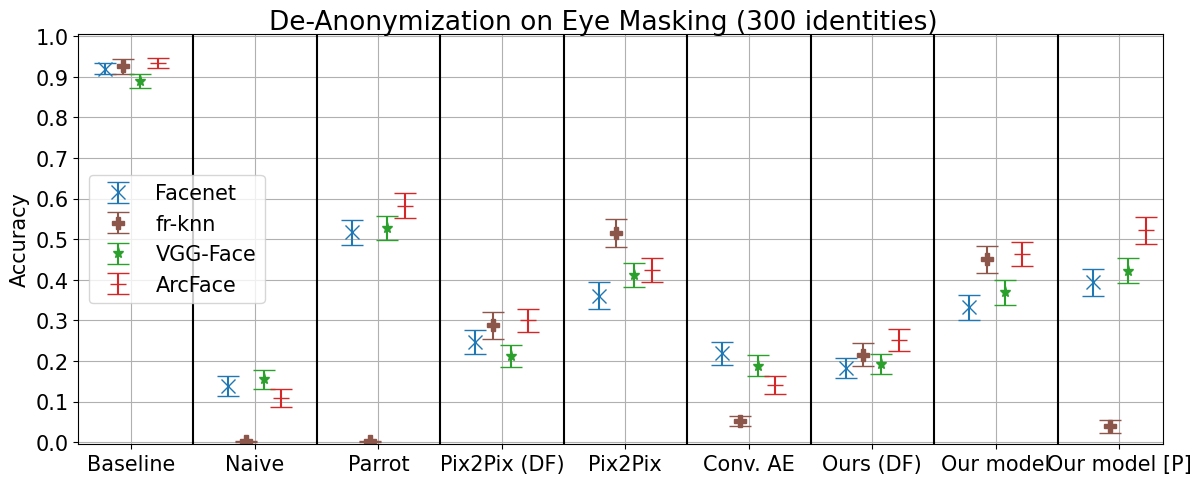}
    \caption{Recognition accuracy for Eye Masking, given for baseline, naive, parrot, de-anon. via Pix2Pix trained on DigiFace-1M, via Pix2Pix, via our model without linear layer, via our model trained on DigiFace-1M, via our model; on 300 identities of CelebA.}
    \label{fig:res-eyemask}
\end{figure}

\begin{figure}[h!]
    \includegraphics[width=\linewidth]{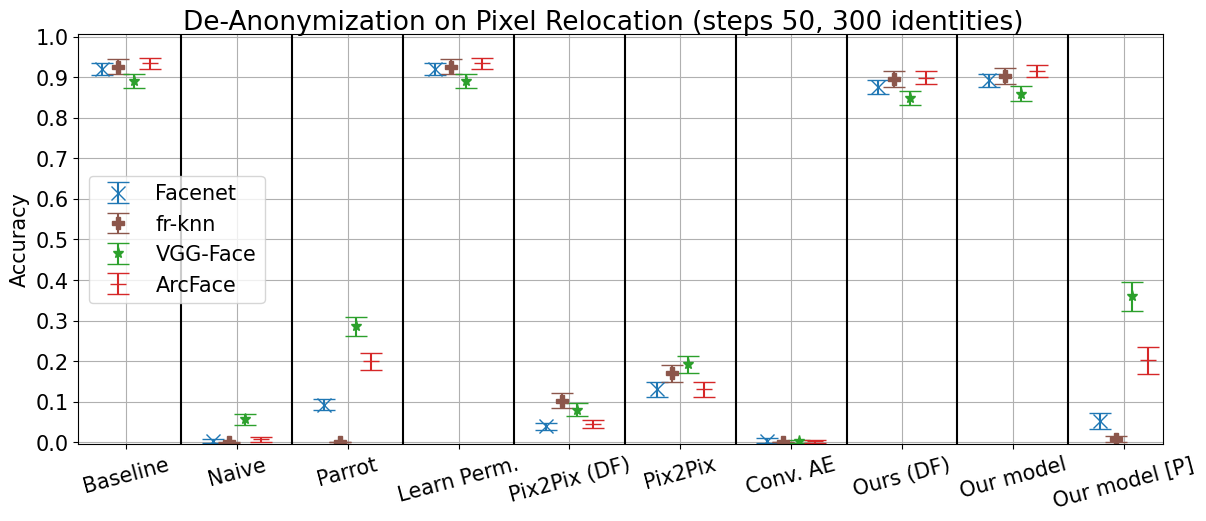}
    \caption{Recognition accuracy for Pixel Relocation, given for baseline, naive, parrot, de-anon. via Learn Permutation, via Pix2Pix trained on DigiFace-1M, via Pix2Pix, via our model without linear layer, via our model trained on DigiFace-1M, via our model; on 300 identities of CelebA.}
    \label{fig:res-pxlreloc}
\end{figure}

\begin{figure}[h!]
    \includegraphics[width=\linewidth]{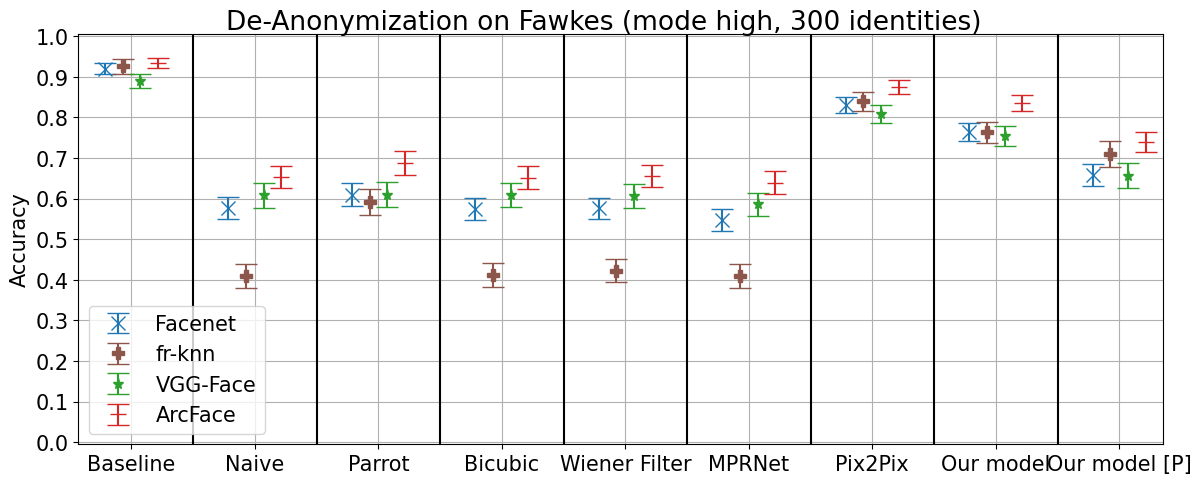}
    \caption{Recognition accuracy for Fawkes, given for baseline, naive, parrot, de-anon. via bicubic interpolation, via Wiener Filter, via MPRNet (Denoising), via Pix2Pix, via our model; on 300 identities of CelebA.}
    \label{fig:res-fawkeshigh}
\end{figure}

\begin{figure}[h!]
    \includegraphics[width=\linewidth]{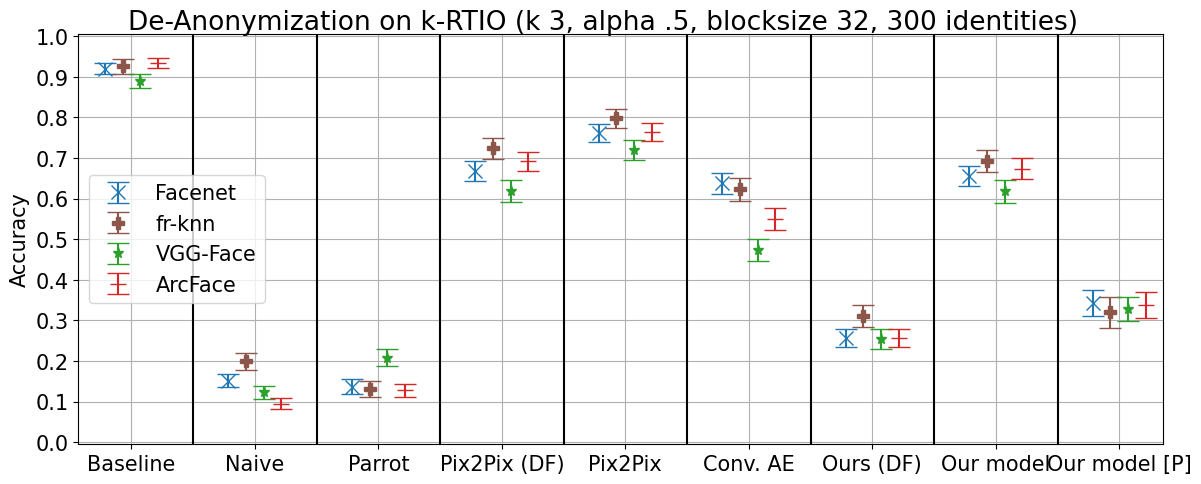}
    \caption{Recognition accuracy for k-RTIO, given for baseline, naive, parrot, de-anon. via Pix2Pix trained on DigiFace-1M, via Pix2Pix, via our model without linear layer, via our model trained on DigiFace-1M, via our model; on 300 identities of CelebA.}
    \label{fig:res-krtio}
\end{figure}

\begin{figure}[h!]
    \includegraphics[width=\linewidth]{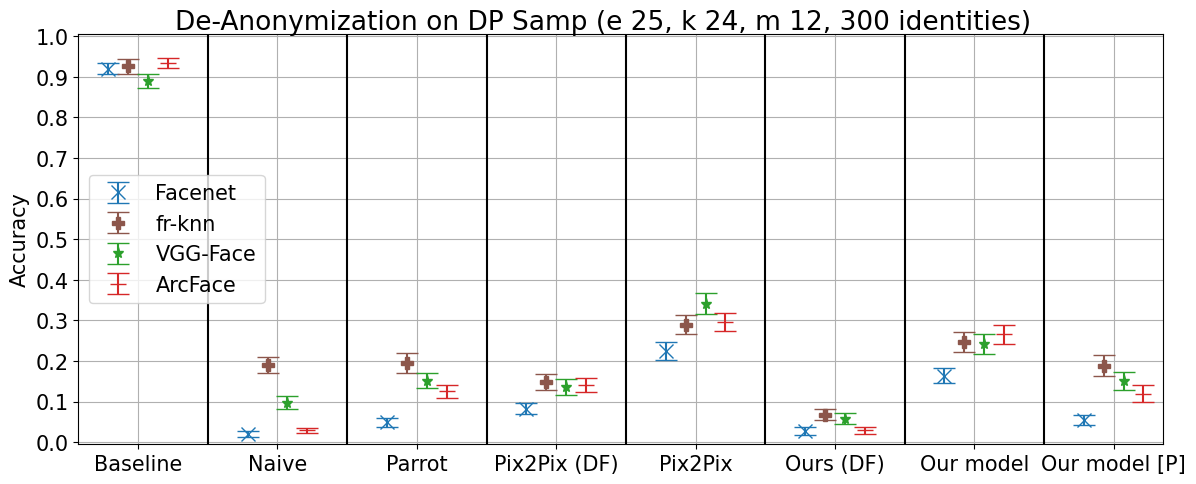}
    \caption{Recognition accuracy for DP Samp, given for baseline, naive, parrot, de-anon. via Pix2Pix trained on DigiFace-1M, via Pix2Pix, via our model trained on DigiFace-1M, via our model; on 300 identities of CelebA.}
    \label{fig:res-dpsamp}
\end{figure}

\begin{figure}[h!]
    \includegraphics[width=\linewidth]{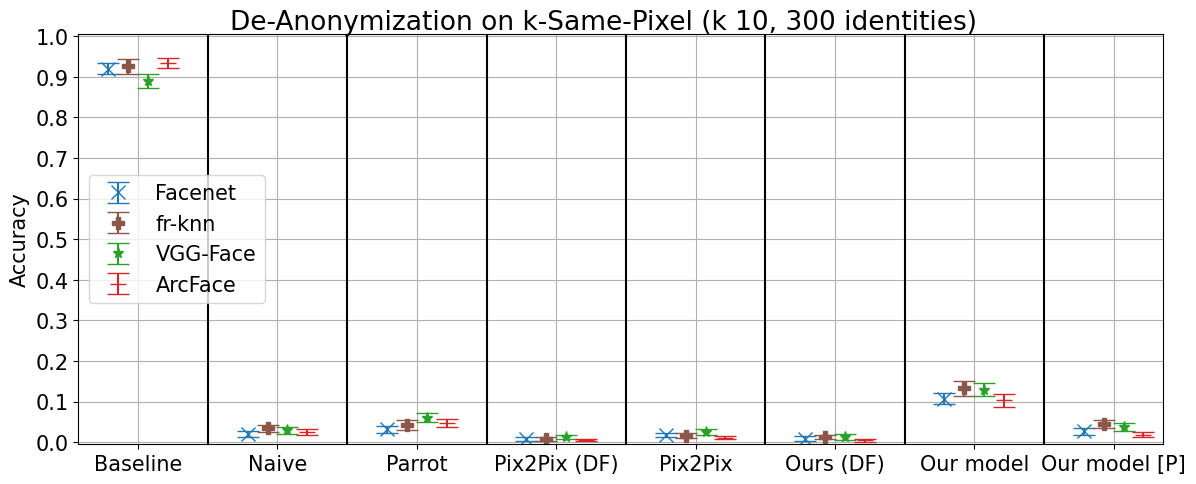}
    \caption{Recognition accuracy for \textit{k}-Same-Pixel, given for baseline, naive, parrot, de-anon. via Pix2Pix trained on DigiFace-1M, via Pix2Pix, via our model trained on DigiFace-1M, via our model; on 300 identities of CelebA.}
    \label{fig:res-ksamepixel}
\end{figure}

\begin{figure}[h!]
    \includegraphics[width=\linewidth]{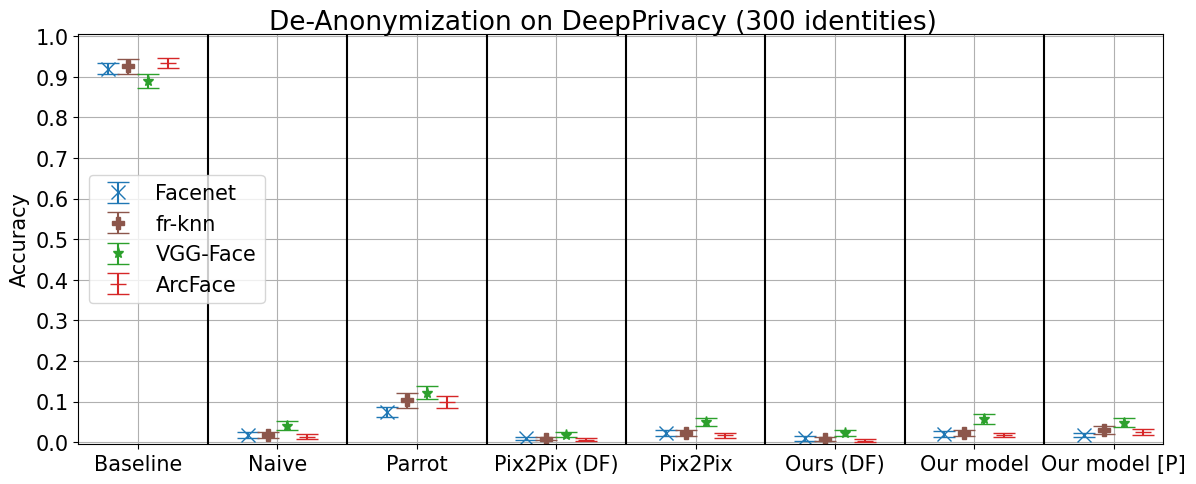}
    \caption{Recognition accuracy for DeepPrivacy, given for baseline, naive, parrot, de-anon. via Pix2Pix trained on DigiFace-1M, via Pix2Pix, via our model trained on DigiFace-1M, via our model; on 300 identities of CelebA.}
    \label{fig:res-deepprivacy}
\end{figure}

\begin{figure}[h!]
    \includegraphics[width=\linewidth]{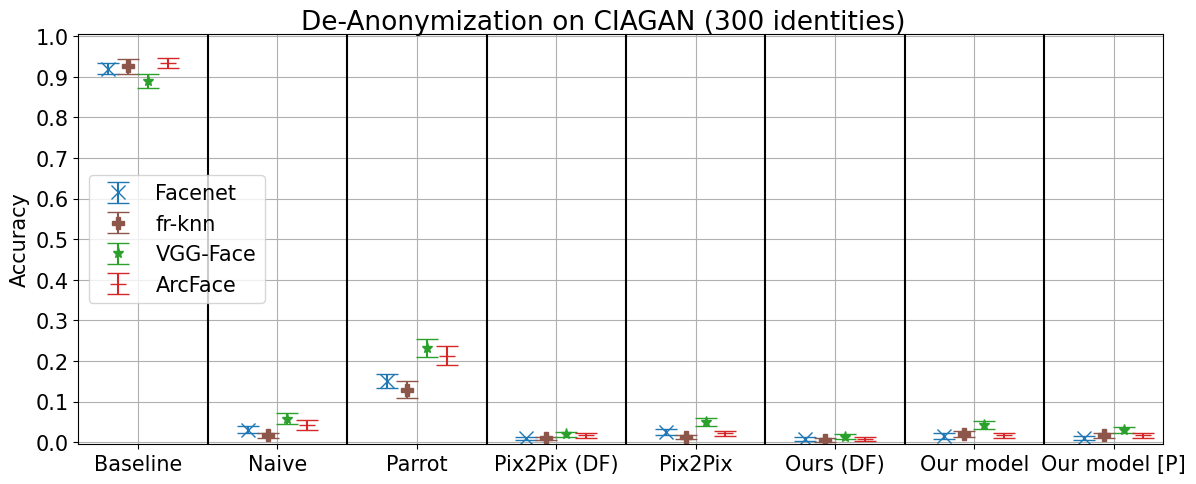}
    \caption{Recognition accuracy for CIAGAN, given for baseline, naive, parrot, de-anon. via Pix2Pix trained on DigiFace-1M, via Pix2Pix, via our model trained on DigiFace-1M, via our model; on 300 identities of CelebA.}
    \label{fig:res-ciagan}
\end{figure}

\begin{figure*}[h!]
    \includegraphics[width=\textwidth]{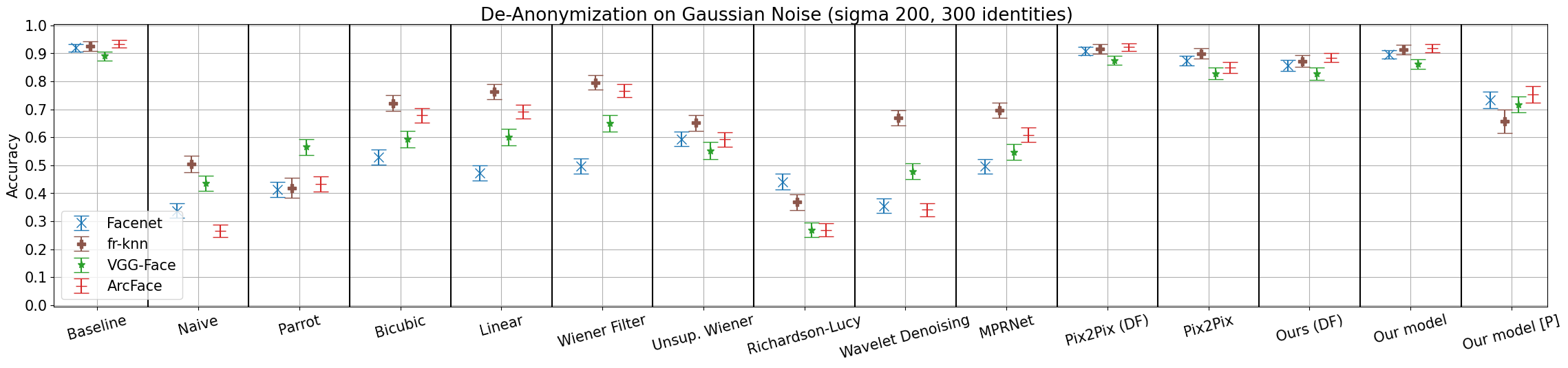}
    \caption{Recognition accuracy for Gaussian Noise, given for baseline, naive, parrot, de-anon. via bicubic interpolation, via linear interpolation, via Wiener Filter, via unsupervised Wiener Filter, via Richardson-Lucy interpolation, via Wavelet Denoising, via MPRNet (Denoising), via Pix2Pix trained on DigiFace-1M, via Pix2Pix, via our model trained on DigiFace-1M, via our model; on 300 identities of CelebA.}
    \label{fig:res-gaussnoise}
\end{figure*}

\begin{figure*}[h!]
    \includegraphics[width=\textwidth]{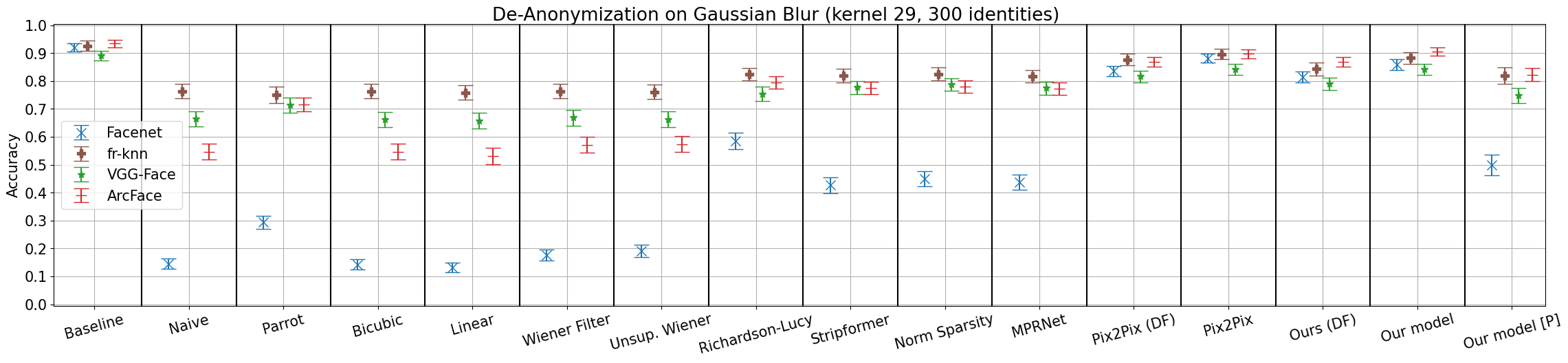}
    \caption{Recognition accuracy for Gaussian Blur, given for baseline, naive, parrot, de-anon. via bicubic interpolation, via linear interpolation, via Wiener Filter, via unsupervised Wiener Filter, via Richardson-Lucy interpolation, via Stripformer, via Norm Sparsity, via MPRNet (Deblurring), via Pix2Pix trained on DigiFace-1M, via Pix2Pix, via our model trained on DigiFace-1M, via our model; on 300 identities of CelebA.}
    \label{fig:res-gaussblur13}
\end{figure*}

\begin{figure*}[h!]
    \includegraphics[width=\textwidth]{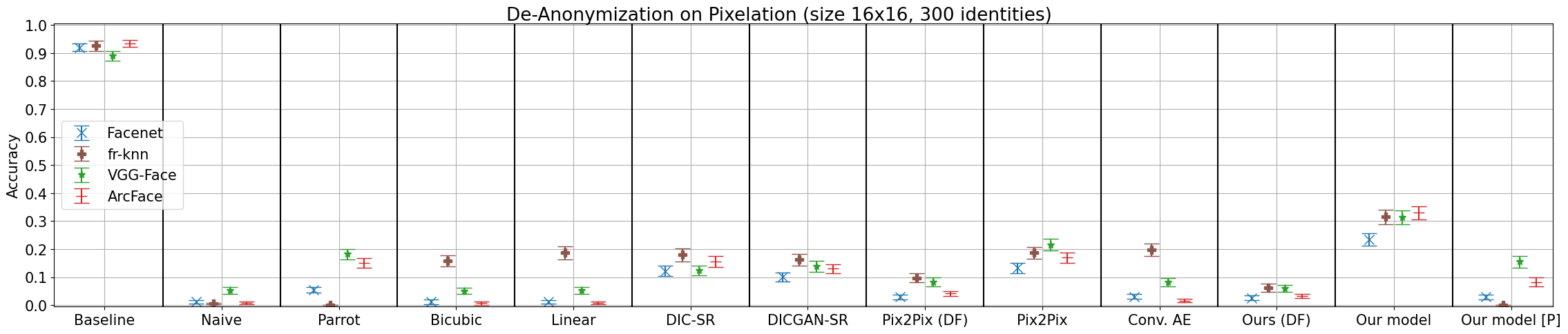}
    \caption{Recognition accuracy for Pixelation, given for baseline, naive, parrot, de-anon. via bicubic interpolation, via linear interpolation, via DIC-SR, via DICGAN-SR, via Pix2Pix trained on DigiFace-1M, via Pix2Pix, via our model without linear layer, via our model trained on DigiFace-1M, via our model; on 300 identities of CelebA.}
    \label{fig:res-pixelate16}
\end{figure*}

\begin{figure*}[h!]
    \includegraphics[width=\textwidth]{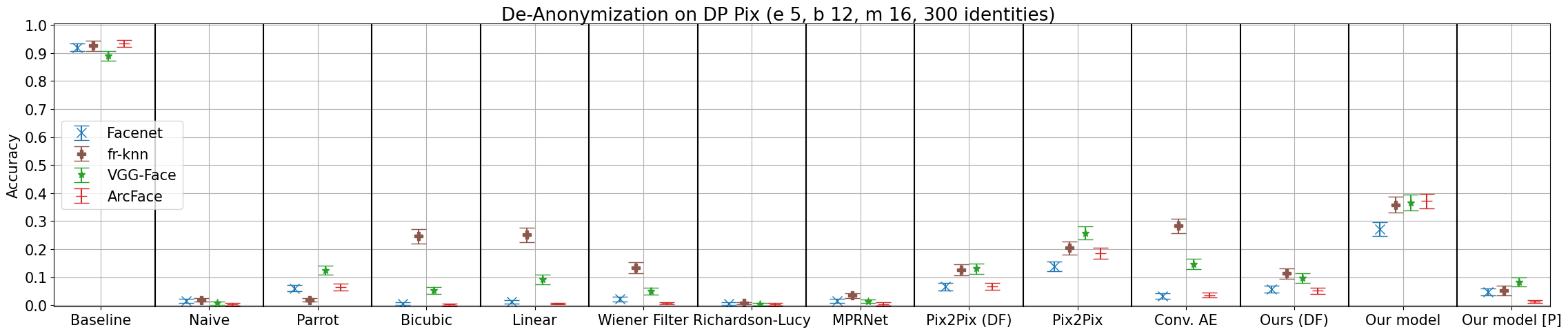}
    \caption{Recognition accuracy for DP Pix, given for baseline, naive, parrot, de-anon. via bicubic interpolation, via linear interpolation, via Wiener Filter, via Richardson-Lucy interpolation, via MPRNet (Denoising), via Pix2Pix trained on DigiFace-1M, via Pix2Pix, via our model without linear layer, via our model trained on DigiFace-1M, via our model; on 300 identities of CelebA.}
    \label{fig:res-dppix}
\end{figure*}

\begin{figure*}[h!]
    \includegraphics[width=\textwidth]{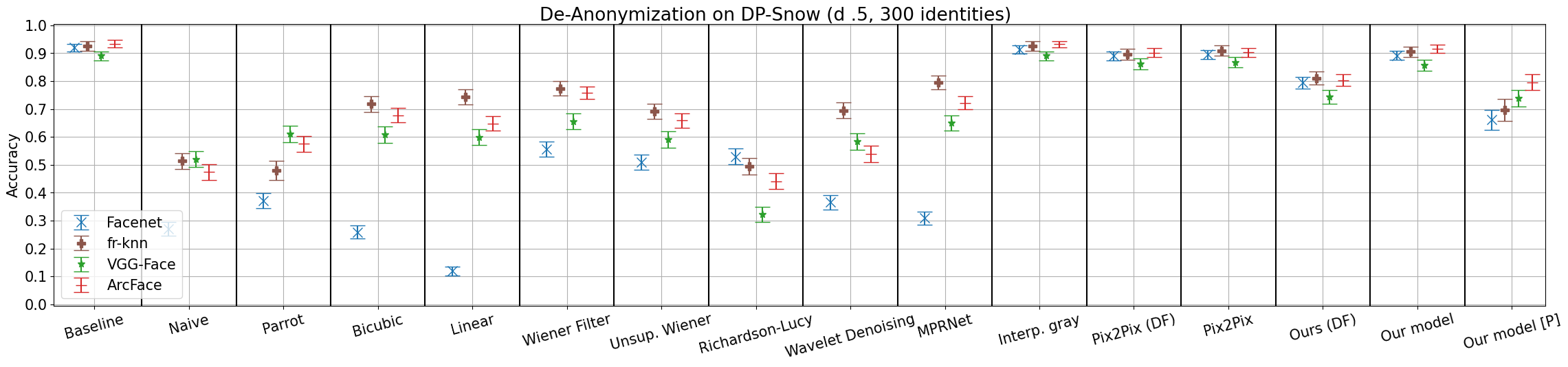}
    \caption{Recognition accuracy for DP Snow, given for baseline, naive, parrot, de-anon. via bicubic interpolation, via linear interpolation, via Wiener Filter, via unsupervised Wiener Filter, via Richardson-Lucy interpolation, via Wavelet Denoising, via MPRNet (Denoising), via Interpolate Gray, via Pix2Pix trained on DigiFace-1M, via Pix2Pix, via our model trained on DigiFace-1M, via our model; on 300 identities of CelebA.}
    \label{fig:res-dpsnow}
\end{figure*}

\begin{figure*}[h!]
    \includegraphics[width=\textwidth]{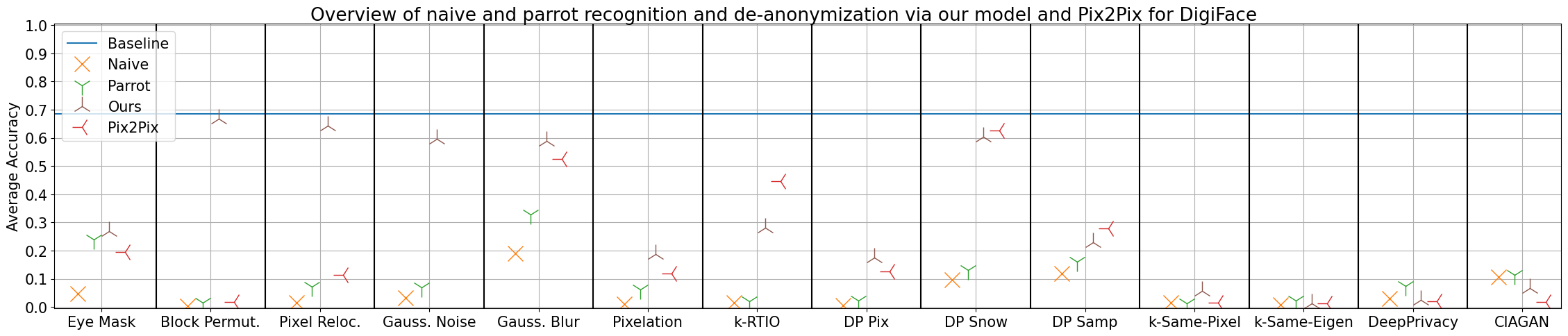}
    \caption{Average recognition accuracy for every anonymization method for baseline, naive, parrot, de-anonymized via our model, and de-anonymized via Pix2Pix; on 300 identities on DigiFace-1M.}
    \label{fig:res-overview-digiface}
\end{figure*}

This section includes further results from the experiments we performed in the context of this work.
For an overview over all anonymizations when using the DigiFace-1M data set, see Figure~\ref{fig:res-overview-digiface}.
See Figure~\ref{fig:res-cmp-models-p2p} for the results of the experiments in which Pix2Pix is trained with other anonymizations than it is tested with.
For the results for specific anonymizations, see Figure~\ref{fig:res-eyemask} for Eye Masking,
Figure~\ref{fig:res-pxlreloc} for Pixel Relocation,
Figure~\ref{fig:res-fawkeshigh} for Fawkes,
Figure~\ref{fig:res-krtio} for k-RTIO,
Figure~\ref{fig:res-dpsamp} for DP Samp,
Figure~\ref{fig:res-ksamepixel} for \textit{k}-Same-Pixel,
Figure~\ref{fig:res-deepprivacy} for DeepPrivacy,
Figure~\ref{fig:res-ciagan} for CIAGAN,
Figure~\ref{fig:res-gaussnoise} for Gaussian Noise,
Figure~\ref{fig:res-gaussblur13} for Gaussian Blur,
Figure~\ref{fig:res-pixelate16} for Pixelation,
Figure~\ref{fig:res-dppix} for DP Pix and
Figure~\ref{fig:res-dpsnow} for DP Snow.

For all of these plots, "(DF)" refers to the model being trained on the DigiFace-1M data set while being tested on CelebA and "[P]" refers to parrot which means that the enrollment data set was anonymized instead of clear.




\end{document}